# Practical Demonstrations of FR3-Band Thin-Film Lithium Niobate Acoustic Filter Design

Taran Anusorn, *Student Member, IEEE*, Omar Barrera, *Student Member, IEEE*, Jack Kramer, *Student Member, IEEE,* Ian Anderson, *Student Member, IEEE,* Ziqian Yao, *Student Member, IEEE*, Vakhtang Chulukhadze, *Student Member, IEEE*, and Ruochen Lu, *Senior Member, IEEE*

***Abstract*—This article presents an approach to control the operating frequency and fractional bandwidth (FBW) of miniature acoustic filters in thin-film lithium niobate (TFLN). More specifically, we used first-order antisymmetric (A1) mode lateral-field-excited bulk acoustic wave resonators (XBARs) to achieve efficient operation at 20.5 GHz. Our technique leverages the thickness-dependent resonant frequency of A1 XBARs, combined with the in-plane anisotropic properties of 128° Y-cut TFLN, to customize filter characteristics. The implemented three-element ladder filter prototype achieves an insertion loss (IL) of only 1.79 dB and a controlled 3-dB FBW of 8.58% at 20.5 GHz, with an out-of-band (OoB) rejection greater than 14.9 dB across the entire FR3 band, while featuring a compact footprint of 0.90 × 0.74 mm$^2$. Moreover, an eight-element filter prototype shows an IL of 3.80 dB, an FBW of 6.12% at 22.0 GHz, and a high OoB rejection of 22.97 dB, demonstrating the potential for expanding to higher-order filters. As frequency allocation requirements become more stringent in future FR3 bands, our technique showcases promising capability in enabling compact and monolithic filter banks toward next-generation acoustic filters for 6G and beyond.**

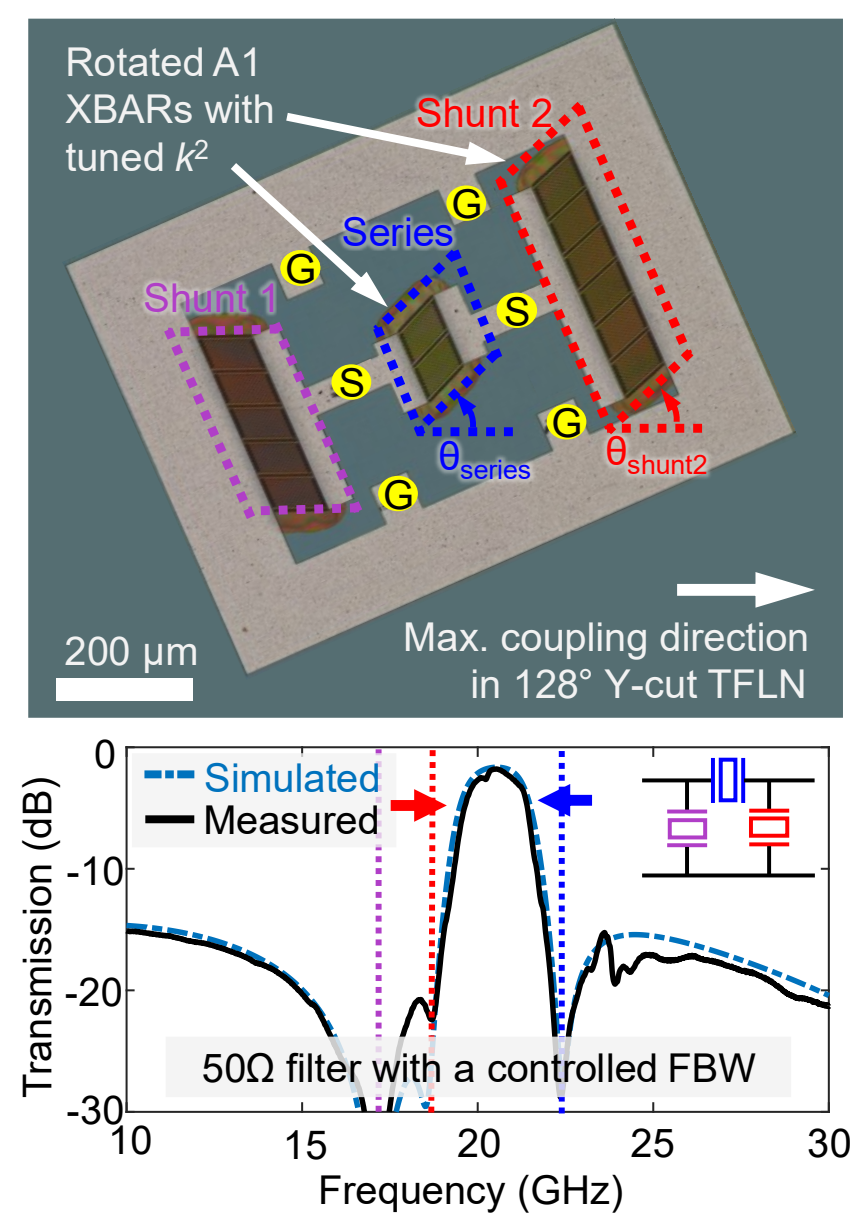

## I. Introduction

FREQUENCY RANGE 3 (FR3), spanning from 7.125 GHz to 24.25 GHz, is emerging as a key spectrum solution for 6G and beyond [1]–[3]. It offers a favorable balance between spectral efficiency, coverage, and deployment cost [2]–[5]. Despite not being finalized yet, the bandwidth (BW) of the FR3 frequency bands is expected to vary from 100 MHz to over 2.5 GHz [6], [7], for the coexistence of multiple standards and services within this spectrum. Thus, precise bandpass filter (BPF) design is essential to minimize interference. While integrated passive device (IPD) filters offer a chip-scale footprint, their moderate roll-off limits the frequency selectivity [7]. Conventional distributed microwave filters, such as waveguides and microstrip patches, can be designed to meet passband specifications [9], [10], but their footprint is relatively large for handheld devices, intrinsically limited by electromagnetic (EM) wavelengths (e.g., wavelength λ ~ 2 cm at 15 GHz). This fact limits their utilization in commercial mobile applications. Several advanced efforts, such as defected ground structures (DGSs) in a printed circuit board (PCB) [11], [12] and slow-wave structures realized by either substrate integrated waveguide (SIW) in a PCB [13], [14], through-silicon via (TSV) technique [15], through-glass via (TGV) technique [16], [17], or low-temperature co-fired ceramic (LTCC) technology [18], [19], have been presented with smaller footprints. While these methods achieve significant miniaturization to sub-wavelength scales, their further miniaturization is ultimately still limited by EM wavelengths. To overcome such limits, acoustic wave filters—possessing wavelengths four to five orders of magnitude shorter than their EM counterparts [20]—hold great promise in enabling chip-scale microwave microsystems [21].

For the past decades, acoustic filters have been dominant mobile filter solutions below 6 GHz owing to their exceptionally low insertion loss (IL) and wide bandwidth (BW)

This work was supported in part by DARPA COFFEE project and in part by Anandamahidol Foundation Scholarship. This article is an expanded version from the IEEE International Microwave Symposium, June 2025.

The authors are with the Department of Electrical and Computing Engineering, The University of Texas at Austin, Austin, TX 78712 USA (Corresponding author: Taran Anusorn, e-mail: taran.anusorn@utexas.edu).

***Highlights***

- **Monolithic FBW control via 128° Y-cut TFLN via rotating A1 XBARs and slant electrodes to exploit in-plane anisotropy for precise bandwidth design.**
- **3-elemen filter: 20.5 GHz, FBW 8.58%, IL 1.79 dB, OoB >14.9 dB in 0.90×0.74 mm². 8-element filter: 22.0 GHz, FBW 6.12%, IL 3.80 dB, OoB 22.97 dB in 1.48×0.89 mm².**
- **The technique enables compact monolithic filter banks with selectable passbands for FR3/6G mobile and sensing, scalable to higher-order networks.**

within a compact footprint [22]. These filters consist of a network of acoustic resonators that convert EM signals into mechanical vibrations and back via the piezoelectric effect [23]. Below 3 GHz, surface acoustic wave (SAW) resonators are commonly used, where standing waves are confined to the surface of a piezoelectric material [24]. The operating frequency ($f_c$) of SAW filters is primarily determined by the lateral distance between the interdigital electrodes (IDEs) of their resonators. However, this design feature limits the operation at high frequencies, as the IDE pitch widths become exceedingly small, causing resistive loss and fabrication challenges [25]. In contrast, bulk acoustic wave (BAW) resonators, which vibrate throughout their acoustic cavity body [26], offer better scalability by reducing the piezoelectric and metal stack thickness. Recent advances in aluminum nitride (AlN)-based film bulk acoustic resonators (FBARs) have shown significant performance improvements above 10 GHz [27]–[30]. Nevertheless, the limited electrotechnical coupling ($k^2$), with the maximum reported value of only 10.4% [30], in such platforms has hindered their ability to achieve wideband filter implementations.

First-order antisymmetric (A1) Lamb mode lateral-field-excited bulk acoustic wave resonators (XBARs) in thin-film lithium niobate (TFLN) [31]–[33] arise as one promising high-frequency acoustic filter platform [34]. Owing to the high quality factor ($Q$) and notably strong $k^2$ up to 46.4% [35], low-loss acoustic filters beyond 20 GHz with a wide fractional bandwidth (FBW) of up to 19.8% have been realized recently using A1 XBARs in 128° Y-cut TFLN [36]–[38]. Nevertheless, a wide FBW is not always an ideal solution, as 6G FR3 requires a specific FBW to match the allocated band. To meet specific FBW requirements, various tuning techniques have been proposed, including circuit-level techniques by adding discrete inductors and capacitors [23] and material-level engineering to adjust effective $k^2$ [39], though these often complicate fabrication and integration. In [40], FBW customization was achieved in monolithic SAW filters via in-plane resonator rotation on a TFLN/$SiO_2$/SiC substrate, enabling compact filter designs for multiple WiFi bands near 2.5 GHz. Building on this concept, we explore a similar rotational tuning mechanism—but applied to A1 XBARs in 128° Y-cut TFLN—to enable monolithic, high-frequency BAW filter design with enhanced flexibility and minimal complexity.

In this work, we leverage the anisotropic properties of 128° Y-cut TFLN to fine-tune $k^2$ of the A1 XBAR, achievable from 0% up to roughly 47%, by rotating its IDEs away from the maximum coupling direction. This approach, combined with TFLN thickness tuning, allows simultaneous control of resonator properties, namely, resonant frequencies and $k^2$, which translates to $f_c$ and FBW of the filter. To validate the technique, we designed and fabricated A1 XBAR ladder filter prototypes targeting FR3 operation with a controlled FBW below 10%. The three-element prototype, with an FBW of 8.58%, achieves a low IL of 1.79 dB and an out-of-band (OoB) rejection level exceeding 14.92 dB across the entire FR3 band, all while maintaining a compact footprint of only 0.90 × 0.74 mm². Moreover, our eight-element prototype yields an FBW of 6.12% with a better minimum OoB rejection of 22.97 dB at the cost of a slightly higher IL of 3.80 dB, showcasing the potential for higher-order filter design. Compared to previously reported FR3 acoustic filters above 10 GHz [36]–[38], [41]–[44], as well as other advanced filter technologies, our results demonstrate exceptional filter performance and design controllability, opening doors toward a precise and flexible design of next-generation XBAR filters.

This article is an extension of [1], which briefly reported preliminary demonstration of frequency and bandwidth control of acoustic filters. Here we extend upon these initial results, improving on the initial demonstration through more compact resonator designs by combining resonator rotation and slanted electrodes, which are widely utilized by SAW filters to suppress unwanted modes [45], as well as increased filter performance, and detailed design trade-off discussions. In addition, this paper covers a detailed design methodology as well as a thorough analysis of fabricated prototypes, highlighting design considerations, challenges, and guidelines for further improvements of TFLN XBAR acoustic filters. The paper is organized as follows. Section II introduces the principles of acoustic filter design, highlighting important aspects of controlling filter characteristics through the properties of acoustic resonators based on mBVD circuit model. Section III presents a finite element analysis (FEA) of rotating A1 XBAR for $k^2$ tuning. Section IV provides a detailed description of the proposed design approach with filter simulation. Section V presents and discusses the measurement results of our fabricated devices. Finally, Section VI states the conclusion.

## II. Principle of Acoustic Filter Design

Designing acoustic filters begins with the determination of acoustic resonator antiresonance, shown in Fig. 1(a), where series resonance ($f_s$) shows high admittance, while parallel resonance ($f_p$) shows low admittance. This behavior can be captured by the modified Butterworth-Van Dyke (mBVD) model [46]. Following Fig. 1(b), the static capacitance ($C_0$) of a resonator is set by the dimensions and the dielectric stack between IDEs. Meanwhile, the motional branch, consisting of the motional capacitance ($C_m$), motional inductance ($L_m$), and motional resistance ($R_m$). These parameters are defined as:

$$C_m = (8/\pi^2) \cdot C_0\, k^2, \tag{1}$$
$$L_m = (\pi^2/8) \cdot 1/(\omega_s^2 C_0 k^2), \tag{2}$$
$$R_m = (\pi^2/8) \cdot 1/(\omega_s C_0 k^2 Q), \tag{3}$$

where $\omega_s = 2\pi\, f_s$, $Q$ is a quality factor, and $k^2$ represents a coupling coefficient. It is essential to note that the $Q$ value in the model is used solely for fitting purposes and may differ from the actual 3-dB quality factor of each resonant peak.

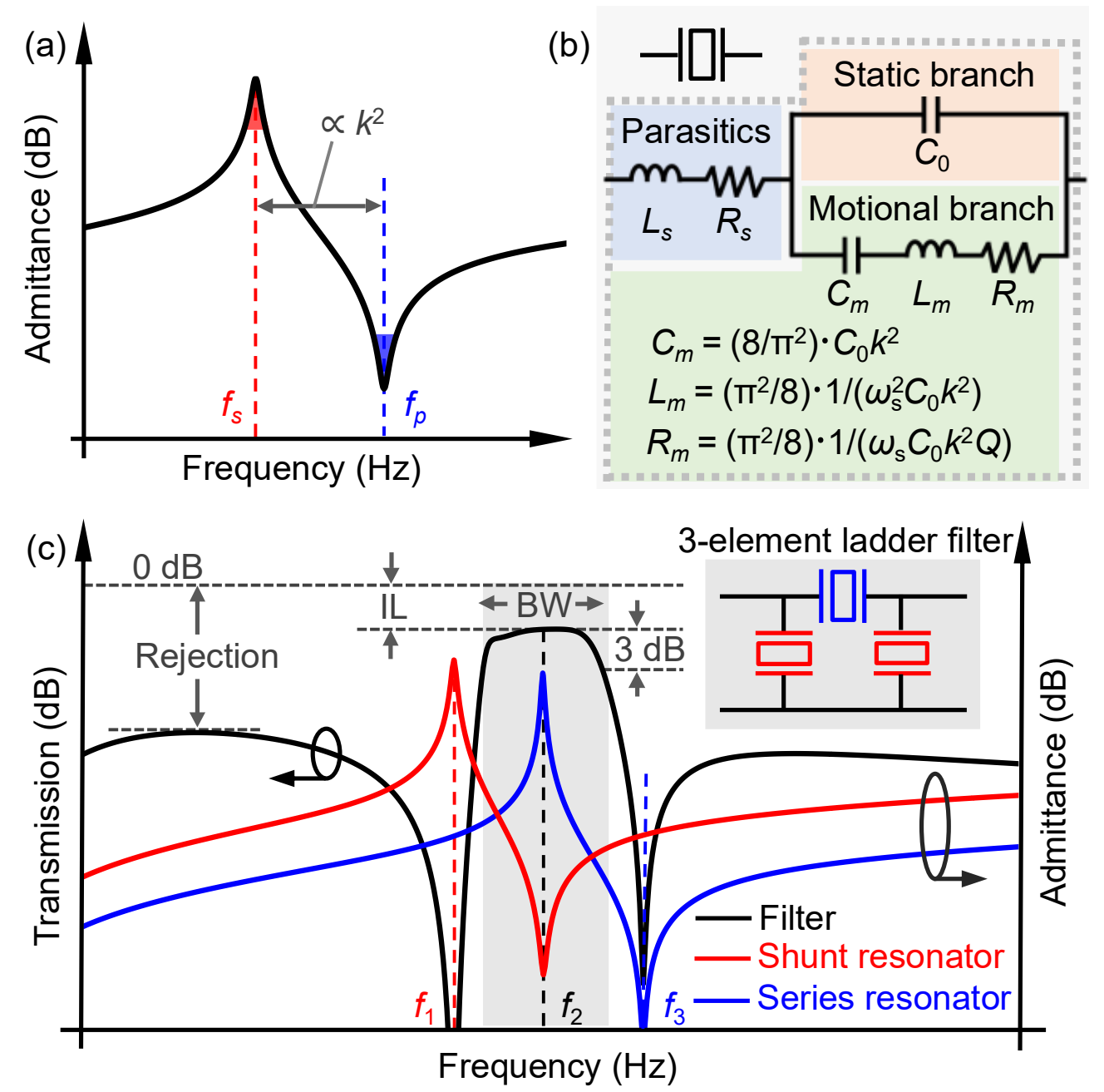

Fig. 1. Important components of acoustic filters. (a) Admittance characteristics over frequency of a generic acoustic resonator. (b) mBVD resonator model. (c) Transmission characteristics of a three-element ladder acoustic filter along with the admittance of the resonators.

In addition to the two branches, the mBVD model incorporates the effects of parasitic elements, such as probing pads, buslines and losses in the resonator, by including a series inductance ($L_s$) and resistance ($R_s$). The impact of $L_s$ and $R_s$ on the acoustic filters will be discussed later in Sections IV and V.

Fig. 1(a) shows the transmission response of a three-element ladder filter, with one series and two shunt resonators, alongside their admittance profiles. At frequency $f_1$, the shunt resonators act as grounds due to high admittance, creating a transmission zero (TZ). At frequency $f_2$, the series resonator exhibits maximum admittance, while the shunt resonators have minimum admittance, enabling effective signal transmission through the filter. The IL of the filters is therefore influenced by $Q$ and $R_s$ of the resonators. Moreover, the filter's $f_c$ can be approximated from $f_s$ of the series resonator. Moving to frequency $f_3$, another TZ is created by $f_p$ of the series resonator. Since the bandwidth (BW) of the filter must lie between these two TZs, determined by $k^2$ of both resonators, FBW can be directly tailored through resonator design.

Systematic approaches for synthesizing acoustic filters have been developed through analytical methods, such as extracted mBVD pole transformation based on Chebyshev [47], [48], or elliptic functions [49], as well as direct bandpass synthesis [50], [51]. Although these methods have proven effective for strategically optimizing filter performance, particularly under stringent requirements or in advanced implementations involving EM structures [52], the intuitive approach described earlier, combined with generic circuit-level optimization of mBVD parameters, has been found sufficient for our work. Nevertheless, the potential directions for improvements based on our results are discussed in Section V.

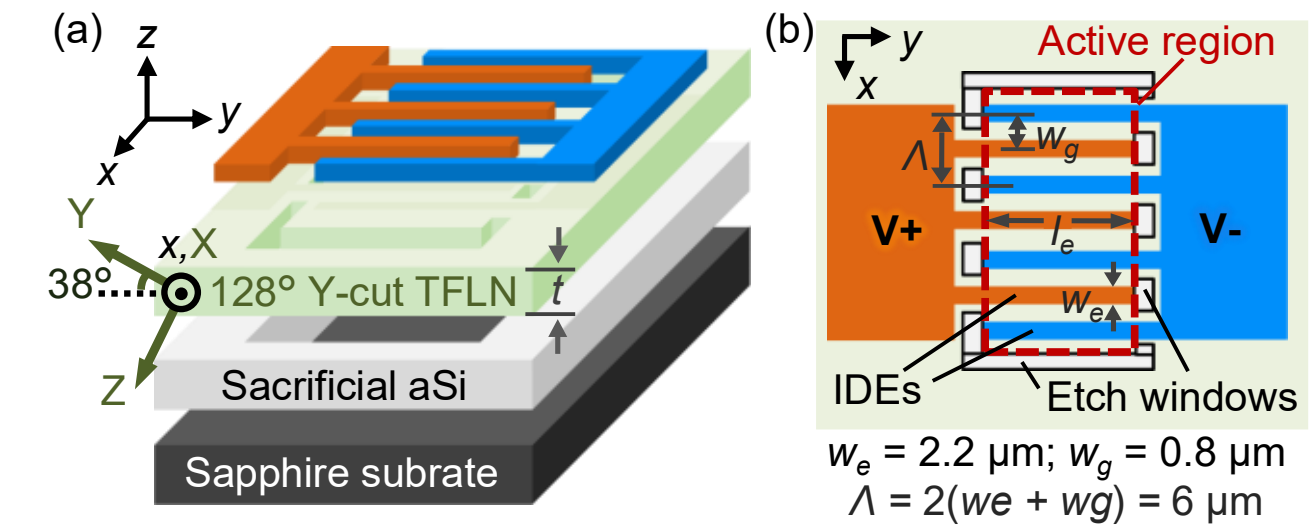

Fig. 2. (a) Exploded and (b) top views of the A1 XBAR in 128° Y-cut TFLN.

## III. First-Order Antisymmetric Mode Resonators

Since the performance of acoustic filters is closely tied to the properties of their constituent resonators, this section provides a comprehensive background on A1 XBARs in 128° Y-cut TFLN in the proposed filter design.

### A. First-Order Antisymmetric Lamb Mode Device

A generic XBAR consists of IDE structures fabricated on top of a suspended thin-film piezoelectric layer (e.g., 128° Y-cut TFLN with film thickness $t$), as shown in Fig. 2(a). The material axes $X$, $Y$, $Z$ will be explained later in this section. To achieve suspension, a sacrificial layer, commonly made of chemically removable materials such as amorphous silicon (aSi), is deposited between the film and a substrate. As shown in Fig. 2(b), etch windows are used to release the thin film and also to define the active region of the resonator. Placing the buslines outside this active region helps prevent the generation of spurious modes and minimizes feedthrough capacitance between the buslines [35]. The dimensions of IDEs, including electrode width $w_e$ and gap width $w_g$, should be chosen to achieve a desirable trade-off between $Q$ and $k^2$ [35].

In this work, we chose $w_e = 2.2$ µm and $w_g = 0.8$ µm, which corresponds to a periodicity of $\Lambda = 2(w_e + w_g) = 6$ µm. When an alternating voltage is applied to the IDEs, an E-field is generated between each IDE pair and laterally excites mechanical vibrations through the piezoelectric effect. Unlike SAW devices, where the wavelength of the laterally excited acoustic wave is determined by the IDE periodicity $\Lambda$, XBARs support a standing BAW, which is dominantly confined within the suspended film by the mechanically free boundaries at both the top and bottom surfaces (i.e., air interfaces). It is worth noting that the width $w_e$ has minimal impact on $k^2$ in XBARs, as long as $w_e$ is significantly larger than the film thickness. Moreover, the IDE length $l_e$ and the number of IDE fingers $N$ only affect $C_0$ of the XBARs. This characteristic is leveraged in our design to fine-tune $C_0$ of each resonator, thereby controlling the resulting filter impedance (details in Section V).

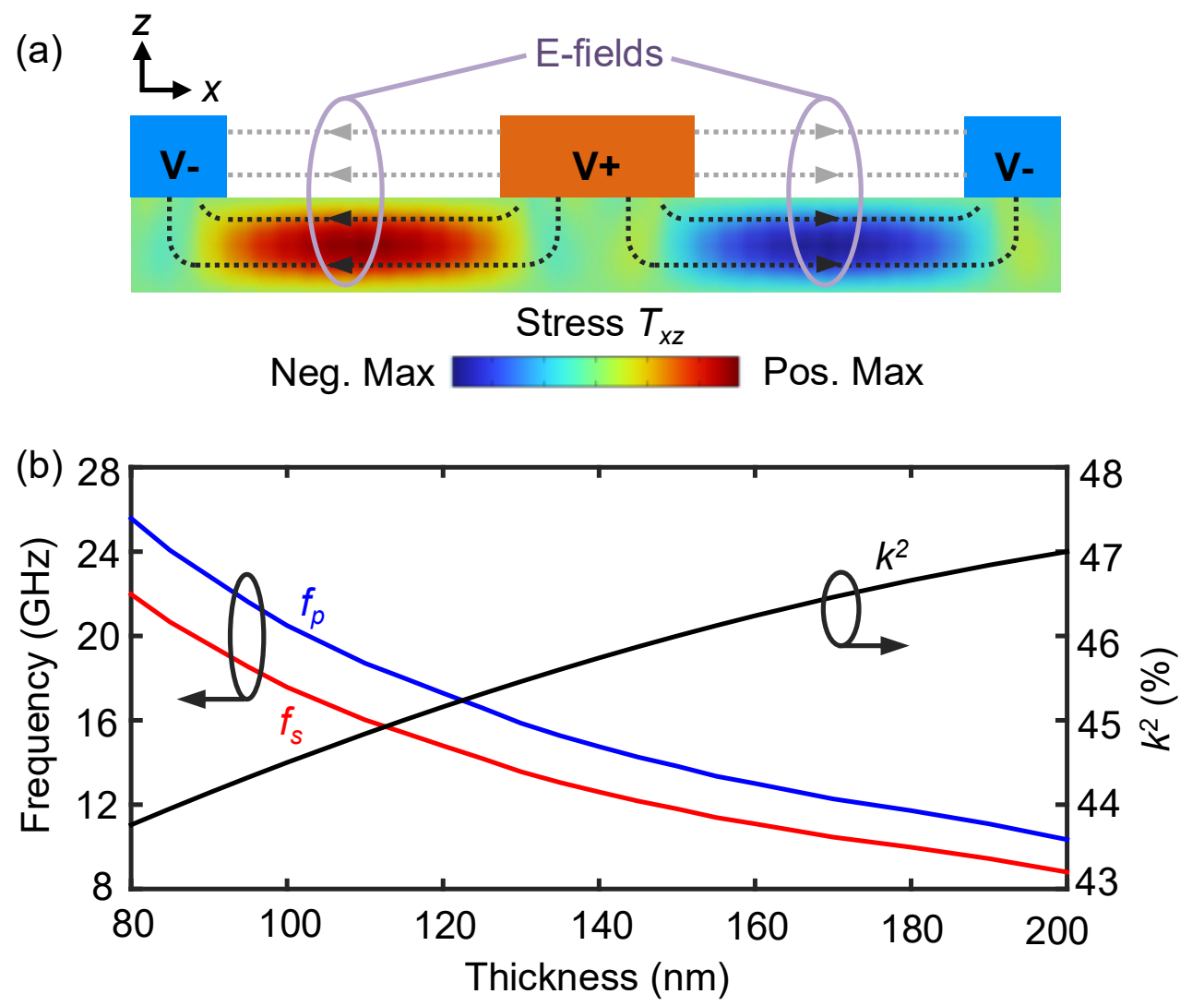

Fig. 3. (a) Simulated stress ($T_{xz}$) mode shape of at the A1 resonance with illustrative lateral-excited E-fields (dimensions are not to scale). (b) Dispersion curves of the A1 resonance at different film thicknesses along with the perceived $k^2$.

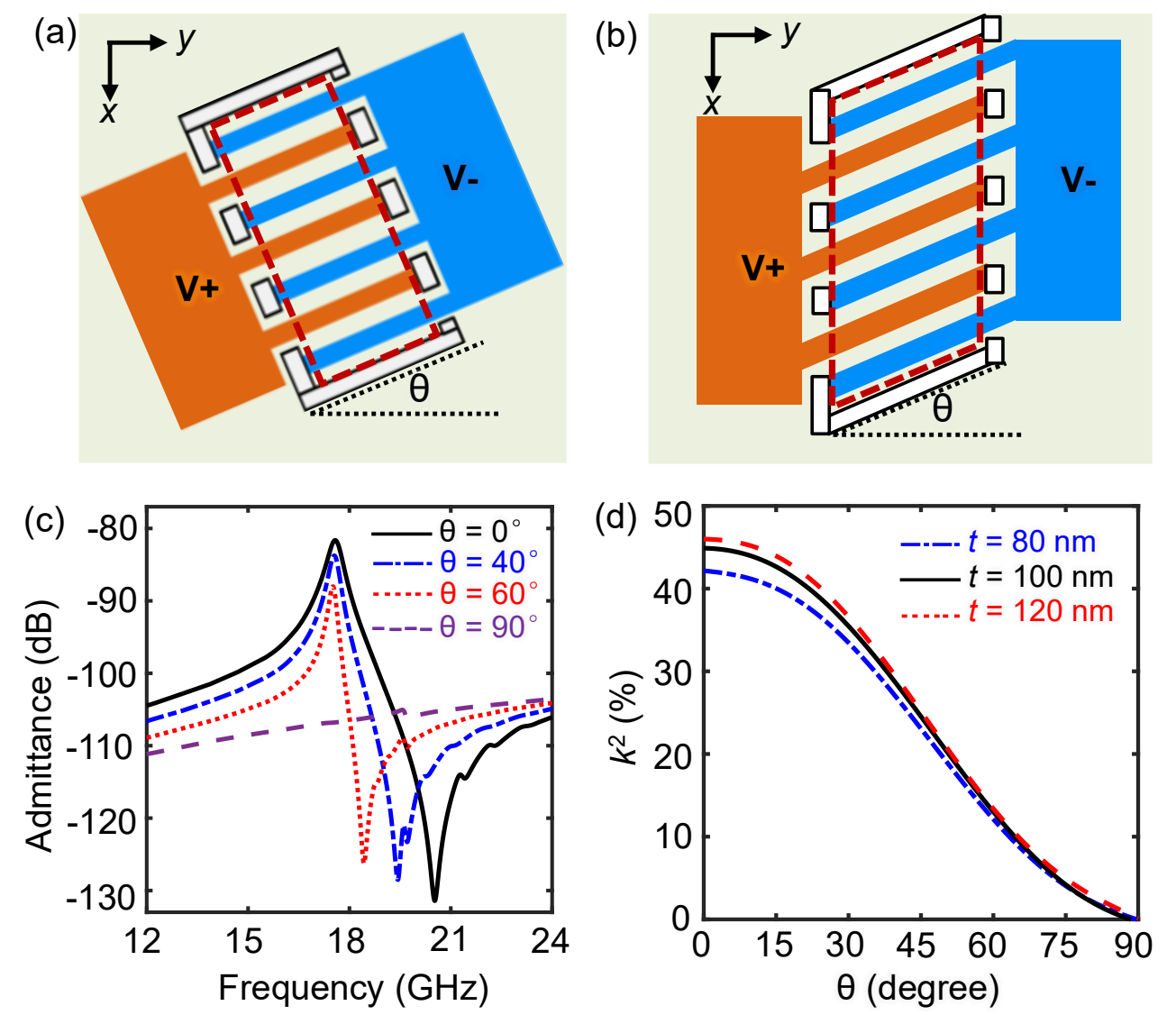

Fig. 4. Lateral E-field reorientation by (a) rotating the entire XBAR and (b) tilting IDEs within the active region. (c) Simulated admittance at different θ with a thickness of 100 nm, indicating the orientation dependency of $k^2$. (d) $k^2$ vs. θ at several TFLN thicknesses ($t$).

Acoustic modes can be classified based on their dominant stress components when the film thickness is significantly smaller than the lateral acoustic wavelength [53]. In 128° Y-cut TFLN, the piezoelectric constant $e_{15}$ is dominant [54], resulting in the strong excitation of the A1 mode characterized by shear stress ($T_{xz}$). To illustrate the thickness dependence of A1 resonances, we performed a parametric study of a unit cell with two pairs of IDEs subjected to alternate voltage inputs in COMSOL FEA. Periodic boundary conditions were assigned to four lateral edges of the unit cell, while its top and bottom were sandwiched by air layers. The simulated A1 mode pattern, showing $T_{xz}$ distribution, is presented in Fig. 3(a). The resulting dispersion curves for $f_s$ and $f_p$, along with the perceived $k^2$ [55]:

$$k^2 = (\pi^2/8) \cdot [(f_p/f_s)^2 - 1], \tag{4}$$

are plotted in Fig. 3(b). This data will serve as the basis for determining resonator thicknesses by film trimming (explained in Section V) according to the filter design methodology.

### B. In-plane Anisotropic Properties of 128° Y-cut TFLN

The anisotropic piezoelectric constants of TFLN can be evaluated by transforming the original material property matrices of the unrotated crystal [56] using the Euler angle rotations. These angles define the orientation of the device coordinate system relative to the material coordinate system, following the $z$–$x$–$z$ rotation convention [57]. For 128° Y-cut TFLN, the corresponding Euler angles are (0°, –38°, 0°) [35]. The corresponding material axes ($X$, $Y$, $Z$), referenced to the local wafer axes ($x$, $y$, $z$), are illustrated in Fig. 2(a). Due to a change in the lateral E-field orientation with respect to the crystal axes, an in-plane rotation of the IDEs on the 128° Y-cut TFLN platform results in a reduction of the effective piezoelectric constant $e_{15}$. This reorientation can be achieved either by rotating the entire resonator or by tilting only the IDEs within the active region, as illustrated in Figs. 4(a) and 4(b), respectively.

The dependency of $e_{15}$ on the rotation has been reported in [58] and a generalized framework for relation between $k^2$ and the orientation angle θ of anisotropic media is presented in [59]. Results from our unit-cell FEA orientation study, plotted in Fig. 4(c), indicate that $f_s$ barely changes with the thickness, whereas $f_p$ exhibits a strong dependency on θ. The perceived $k^2$ variations at different θ and thicknesses, obtained from parametric studies, are shown in Fig. 4(d). This in-plane rotation behavior of $k^2$ introduces an additional design knob for filter design that allows for multiple filters with arbitrary FBWs on the same wafer.

## IV. Filter Design And Simulations

### A. Three-Element Filter Design

Following the principle outlined in Section II, the FBW of a three-element filter can be customized through in-plane rotation of IDEs in 128° Y-cut TFLN using the principle presented in Section III. For example, an FBW-tuned filter shown in Fig. 5(a) can be realized by adjusting $k^2$ of each resonator, as illustrated in Fig. 5(b). It is important to note that $f_s$ and $f_p$ of series and shunt resonators, respectively, are aligned to minimize IL. Since $f_s$ barely changes with IDE orientation, the series resonator can be rotated to the required angle while maintaining a fixed active region thickness. In contrast, both the thickness and orientation of the shunt resonator must be carefully determined. Once the passband of the filter is defined, $C_0$ of each resonator can be tailored to achieve both impedance matching (e.g., 50 Ω) and the desired transmission characteristics. Notably, using a larger shunt resonator (i.e., larger $C_0$) improves OoB rejection but worsens IL.

In addition to FBW control, in-plane orientation enables tuning of TZs to different frequencies. For instance, the shunt resonators in the three-element filter can be designed with different $f_s$ and $k^2$ to introduce an additional TZ, as shown in Fig. 6(a). The added TZ can be strategically positioned to enhance either the selectivity by sharpening the passband roll-off or improve the OoB rejection (defined as a minimum IL within 10BW away from $f_c$) at the lower band, as illustrated in Fig. 6(b). Notably, in the latter case, the passband of the filter becomes slightly higher as the shunt resonator contributing to the close-in rejection is smaller and the added TZ 'stretches' the filter response. This idea can also be applied to series resonators in higher-order ladder filters.

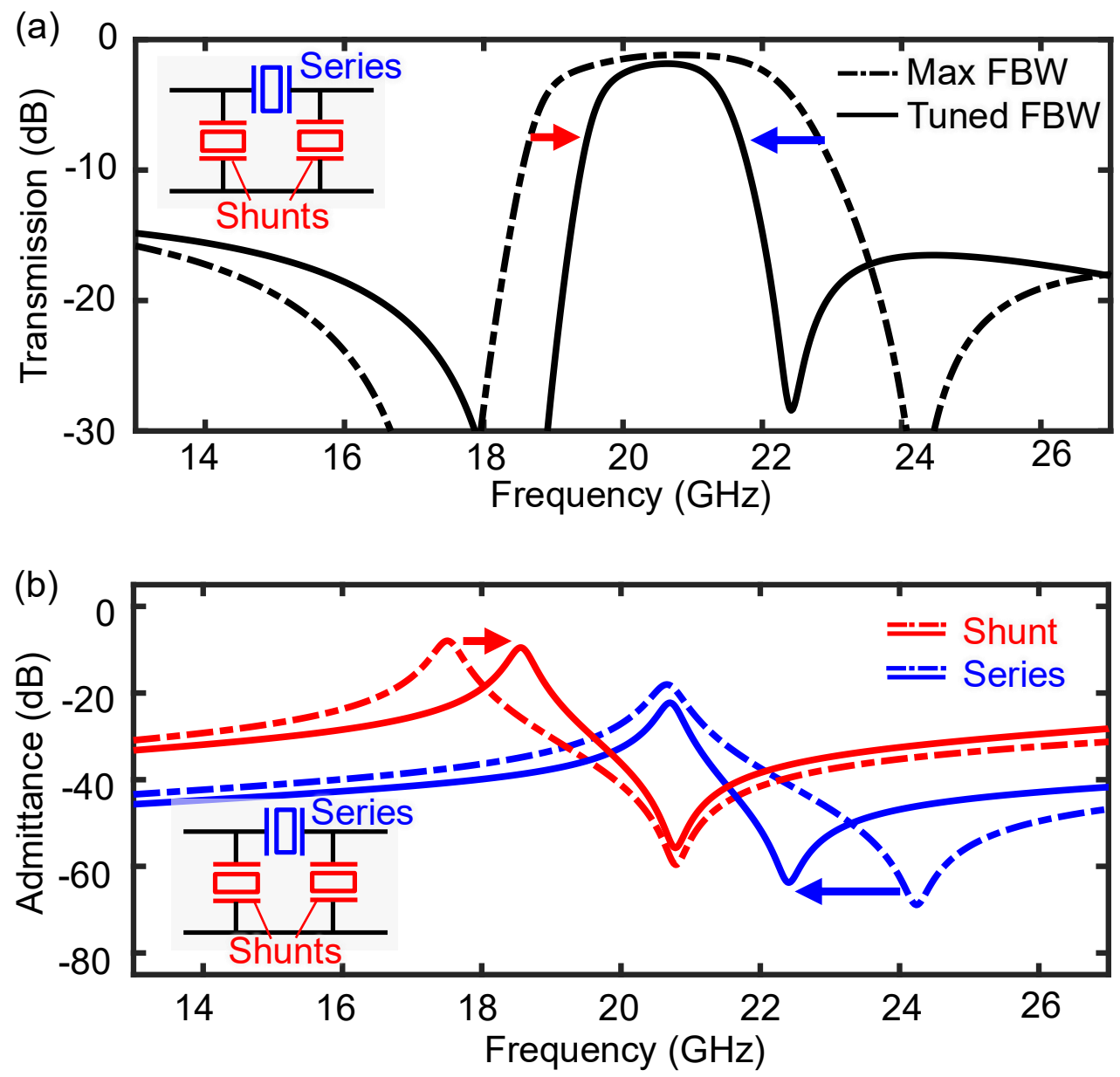

Fig. 5. (a) FBW of the three-element filter is tuned from the maximum achievable FBW by rotating the IDEs of each resonator. (b) $k^2$ tuning in each resonator from the maximum FBW orientation.

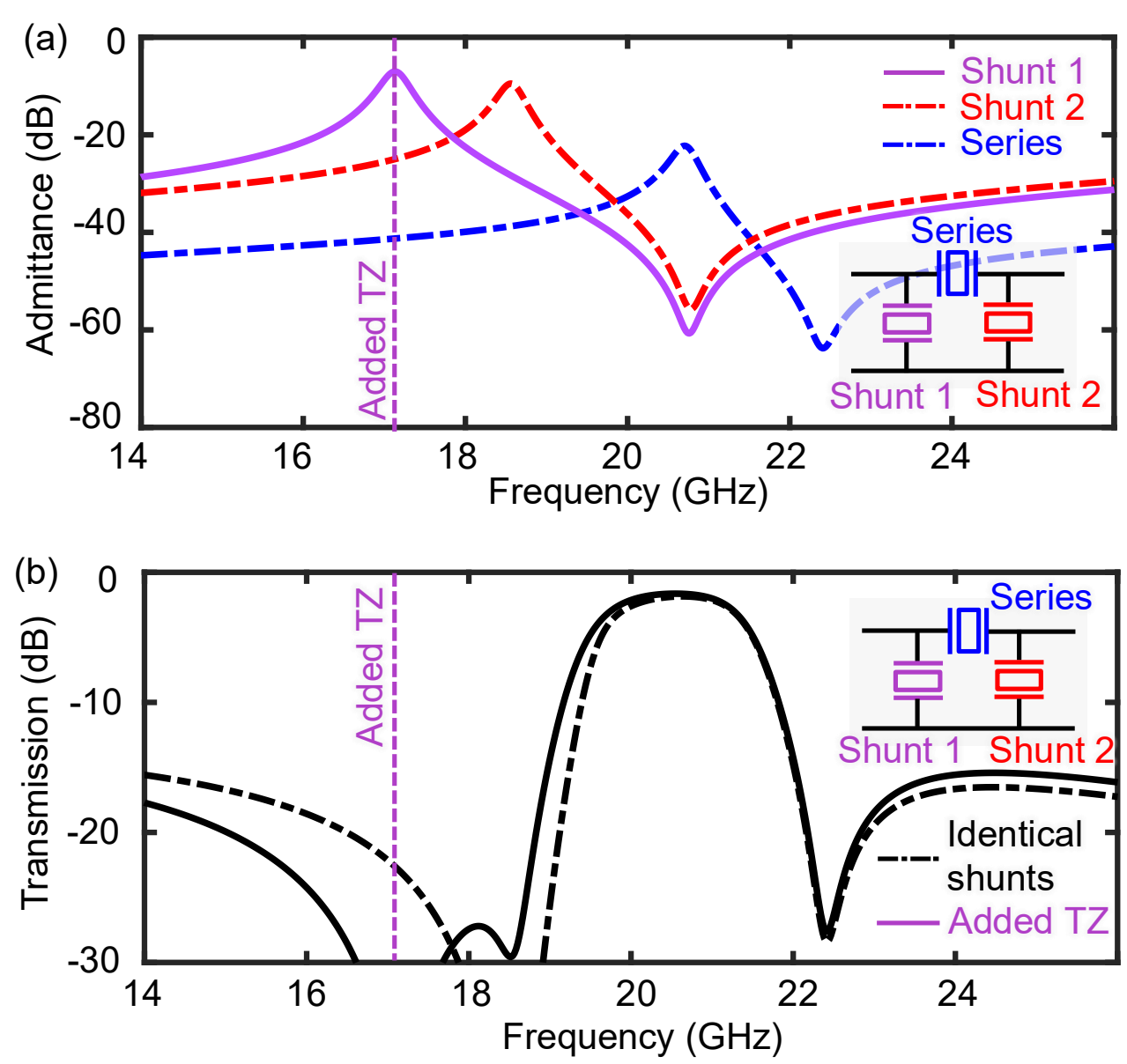

Fig. 6. (a) An additional TZ is introduced by changing the properties of one shunt resonator. (b) OoB rejection is improved by the added TZ.

To demonstrate the proposed design approach, a 50-Ω BPF centered at 20.5 GHz was designed using one series resonator

TABLE I
DESIGN PARAMETERS OF THREE-ELEMENT PROTOTYPE

| Component | mBVD parameters | Device realization |
|---|---|---|
| Series | $f_s$ = 21 GHz | $t$ = 83 nm (2 trims) |
| | $k^2$ = 17% | θ = 50° (φ = 25°) |
| | $C_0$ = 48 fF | $N_e$ = 16 |
| | | $N_g$ = 2 |
| | | $l_e$ = 62 μm |
| | $Q$ = 80, $R_s$ = 3 Ω, $L_s$ = 0.2 nH | |
| Shunt 1 | $f_s$ = 17.9 GHz | $t$ = 99 nm (no trim) |
| | $k^2$ = 42.5% | θ = 0° (φ = –25°) |
| | $C_0$ = 168 fF | $N_e$ = 16 |
| | | $N_g$ = 5 |
| | | $l_e$ = 68 μm |
| | $Q$ = 80, $R_s$ = 1 Ω, $L_s$ = 0.1 nH | |
| Shunt 2 | $f_s$ = 19.1 GHz | $t$ = 92 nm (1 trim) |
| | $k^2$ = 22.6% | θ = 45° (φ = 20°) |
| | $C_0$ = 169 fF | $N_e$ = 17 |
| | | $N_g$ = 6 |
| | | $l_e$ = 63 μm |
| | $Q$ = 80, $R_s$ = 1 Ω, $L_s$ = 0.1 nH | |

$w_e$ = 2.2 μm, $w_g$ = 0.8 μm, and $\Lambda$ = 6 μm.

and two distinct shunt resonators. In our previous study [1], we presented a straightforward method of tuning frequency by rotating the entire resonators. In this work, we extend that concept by combining resonator rotation with IDE tilting to further reduce the overall filter footprint, significantly reducing its size. The targeted FBW of 9.5% was achieved by simultaneously tuning the in-plane orientation, using a global filter rotation angle of α = 25° and an additional IDE tilting angle φ = θ – α, and adjusting the TFLN thickness of each resonator. Note that the axes $x$ and $y$ here indicate the local coordinate of the wafer, not the material. The required $C_0$ for each resonator was realized by calculating the appropriate number of IDE pairs and their finger lengths $l_e$, using the capacitance density obtained from the unit cell simulation. For fabrication purposes, the total number of IDEs was evenly divided into several groups, defined by the number of groups ($N_g$) and the number of IDEs per group ($N_e$), to ensure complete release of the active region while minimizing the risk of structural collapse. The design parameters were optimized to achieve a well-balanced trade-off between IL and OoB rejection. Table I summarizes the optimized design parameters for the proposed filter prototype. Note that the target $f_s$ of each resonator was realized through a series of TFLN thickness trimming steps. The detailed fabrication process will be explained in Section V. It is important to note that $Q$, $R_s$, and $L_s$ of each resonator used in our design are estimated based on our previous studies [36], [37], [43].

## *B. Eight-Element Filter Design*

To further demonstrate the functionality of the proposed design approach, a higher-order filter was developed. Film thickness inspection revealed a slight thickness variation in our TFLN wafer, with the film being approximately 3 nm thinner at the intended fabrication location. To avoid additional fabrication steps, the eight-element filter was designed to reuse

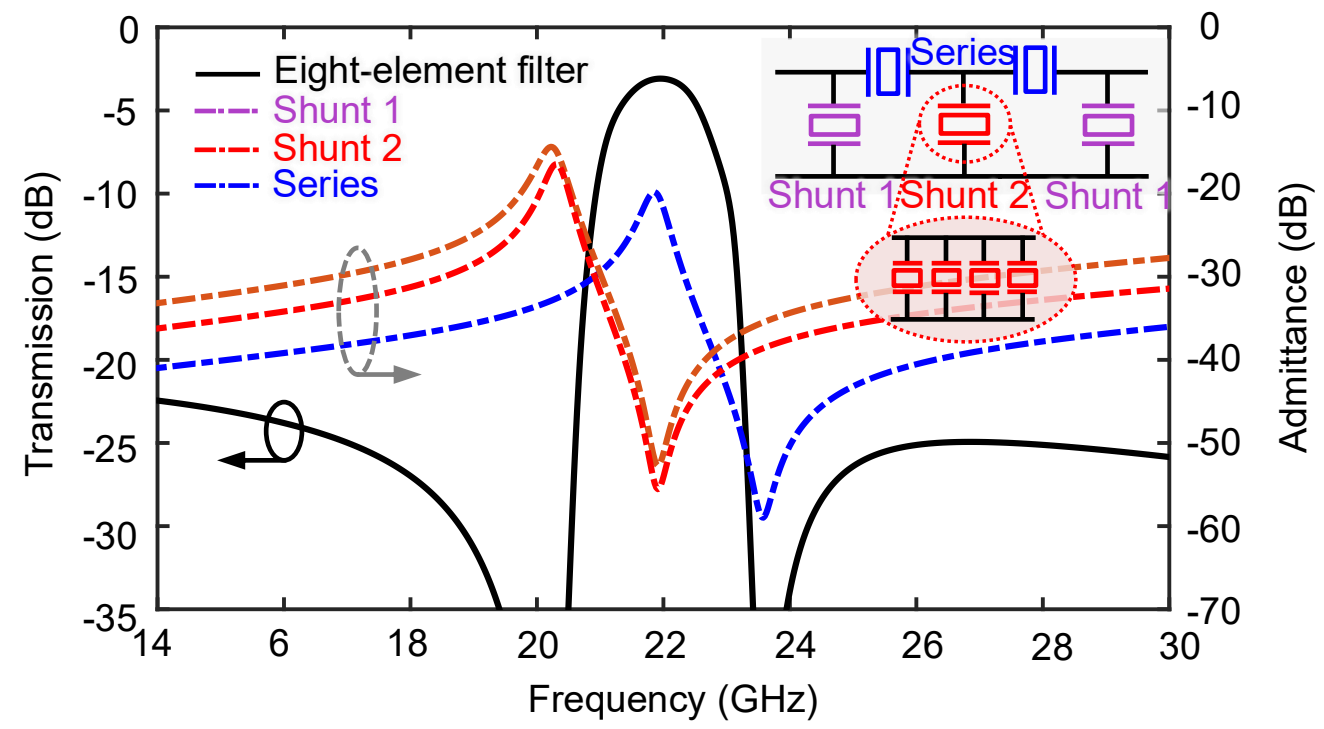

Fig. 7. Frequency responses of the eight-element filter, with an equivalent circuit schematic shown in the inlet, and its constituent resonators.

TABLE II
DESIGN PARAMETERS OF EIGHT-ELEMENT PROTOTYPE

| Component | mBVD parameters | Device realization |
|---|---|---|
| Series | $f_s$ = 22.13 GHz | $t$ = 80 nm (2 trims) |
| | $k^2$ = 16.5% | θ = 50° (φ = 25°) |
| | $C_0$ = 77 fF | $N_e$ = 16 |
| | | $N_g$ = 3 |
| | | $l_e$ = 63 μm |
| | $Q$ = 80, $R_s$ = 3.5 Ω, $L_s$ = 0.1 nH | |
| Shunt 1 | $f_s$ = 20.5 GHz | $t$ = 89 nm (1 trim) |
| | $k^2$ = 17.5% | θ = 50° (φ = 25°) |
| | $C_0$ = 180 fF | $N_e$ = 16 |
| | | $N_g$ = 8 |
| | | $l_e$ = 61 μm |
| | $Q$ = 80, $R_s$ = 2.5 Ω, $L_s$ = 0.05 nH | |
| Shunt 2 | $f_s$ = 20.5 GHz | $t$ = 89 nm (1 trim) |
| | $k^2$ = 17.5% | θ = 50° (φ = 25°) |
| | $C_0$ = 130 fF | $N_e$ = 17 |
| | | $N_g$ = 4* |
| | | $l_e$ = 69 μm |
| | $Q$ = 80, $R_s$ = 2.5 Ω, $L_s$ = 0.0.5 nH | |

$w_e$ = 2.2 μm, $w_g$ = 0.8 μm, and $\Lambda$ = 6 μm.
*Shunt 2 is divided into 4 identical resonators ($C_0$ = 32.5 fF each).

the same trimming depths as those used in the three-element prototype (7 nm and 9 nm for the first and second trims, respectively). As a result, the center frequency $f_c$ of the eight-element prototype is 22.0 GHz. Moreover, the orientation combination was slightly adjusted to achieve an FBW of 6.4% in this case to illustrate the capability of fabricating filters with different FBWs on the same wafer by using the proposed methodology.

To mitigate the increased IL associated with additional series resonators, the design employs two identical series resonators and a single Shunt 2 resonator positioned between them, with two identical Shunt 1 resonators at the input ports. The simulated transmission response of this filter is shown in Fig. 7, alongside that of the lower-order prototype, highlighting a significant improvement in OoB rejection. To accommodate layout wiring and spacing constraints, the Shunt 2 resonator was subdivided into four smaller, identical resonators. The design parameters for the eight-element prototype are summarized in Table II.

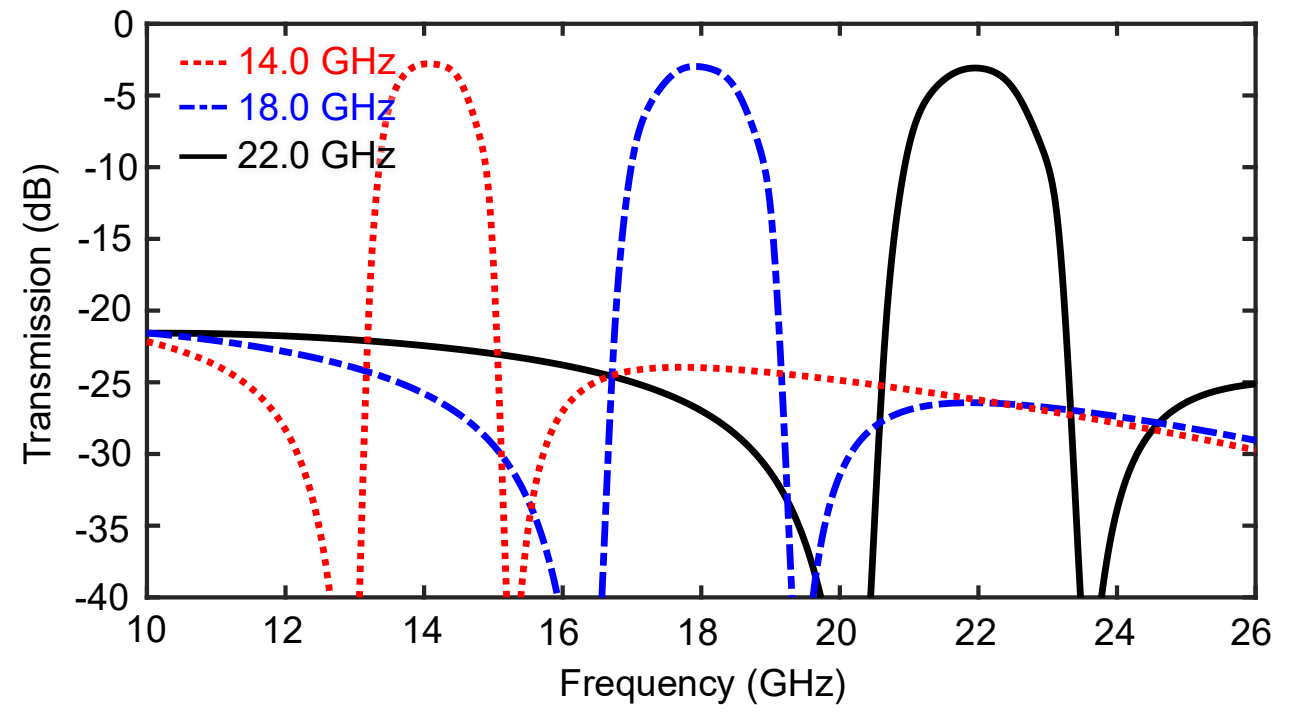

Fig. 8. Simulated frequency-scaled filters with the same layout.

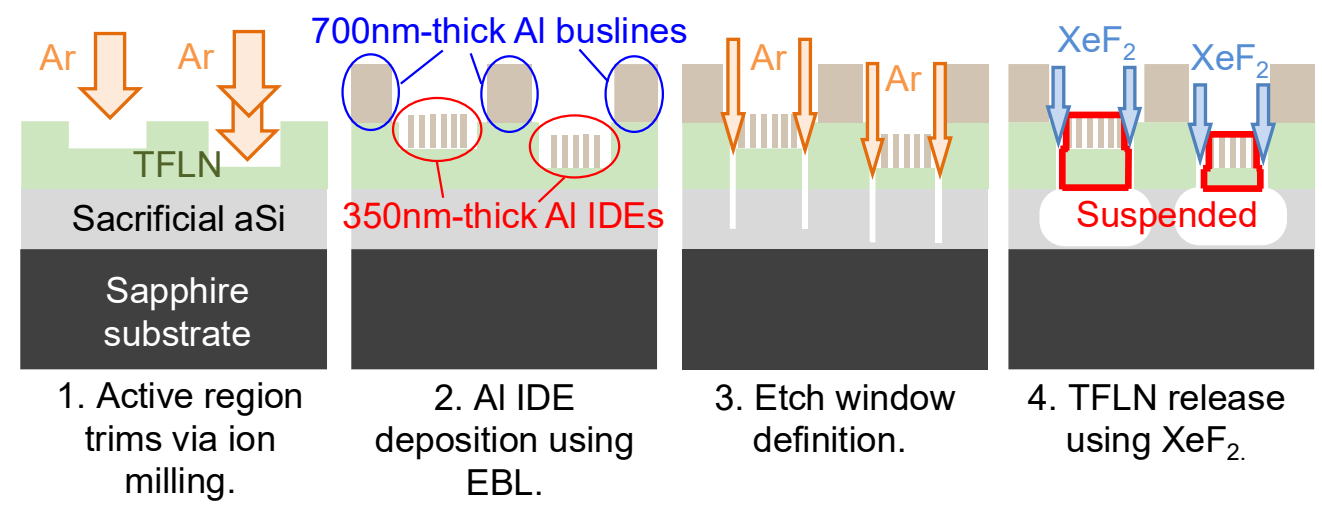

Fig. 9. In-house fabrication process.

## C. Impact of Constant Resonator Impedance

In addition to the capability to scale the operating frequency, our study also reveals that static capacitance $C_0$ is proportional to the TFLN thickness. As a result, the resonator impedance, $Z_{res} \propto 1/(f_c C_0)$, remains constant with variations in thickness. This elegant property enables proportional scaling of the operating frequency band by simply adjusting the base thickness, as illustrated in Fig. 8, while preserving the same filter layout. Although this characteristic lies outside the primary scope of this study, it is worth noting that such behavior is unique to XBAR filters and is not observed in incumbent FBAR and SAW technologies [1]. This distinctive capability opens new avenues for further exploration in acoustic filter design.

# V. FABRICATION RESULTS AND DISCUSSIONS

## A. Device Fabrication

Fig. 9 summarizes the fabrication process utilized in this work. The process begins with a TFLN-on-aSi-on-Sapphire wafer provided by NGK Insulator Ltd. For this demonstration, target thicknesses were realized through two sequential ion milling steps with distinct exposure times: the first trim, removing 7 nm, was applied to both shunt and series resonators; the second trim, removing an additional 9 nm, was applied exclusively to the series resonators. In each step, high-energy ions (e.g., ionized Ar gas) bombard the surface within a vacuum chamber, selectively thinning the TFLN layer and thereby defining the active region thickness of each resonator. The thickness of each trimmed region was measured by Filmetrics optical profilometer, indicating only ±1 nm deviation. Notably, this physical etching process preserves the crystalline quality of the TFLN layer, which is essential for retaining its piezoelectric properties [60]. Due to the minor thickness variations across the wafer for research prototyping purposes, frequency shifts in the devices are expected. However, as discussed in Section IV, such variations primarily result in a uniform shift of the center frequency $f_c$, provided that the trimmed regions undergo consistent thinning.

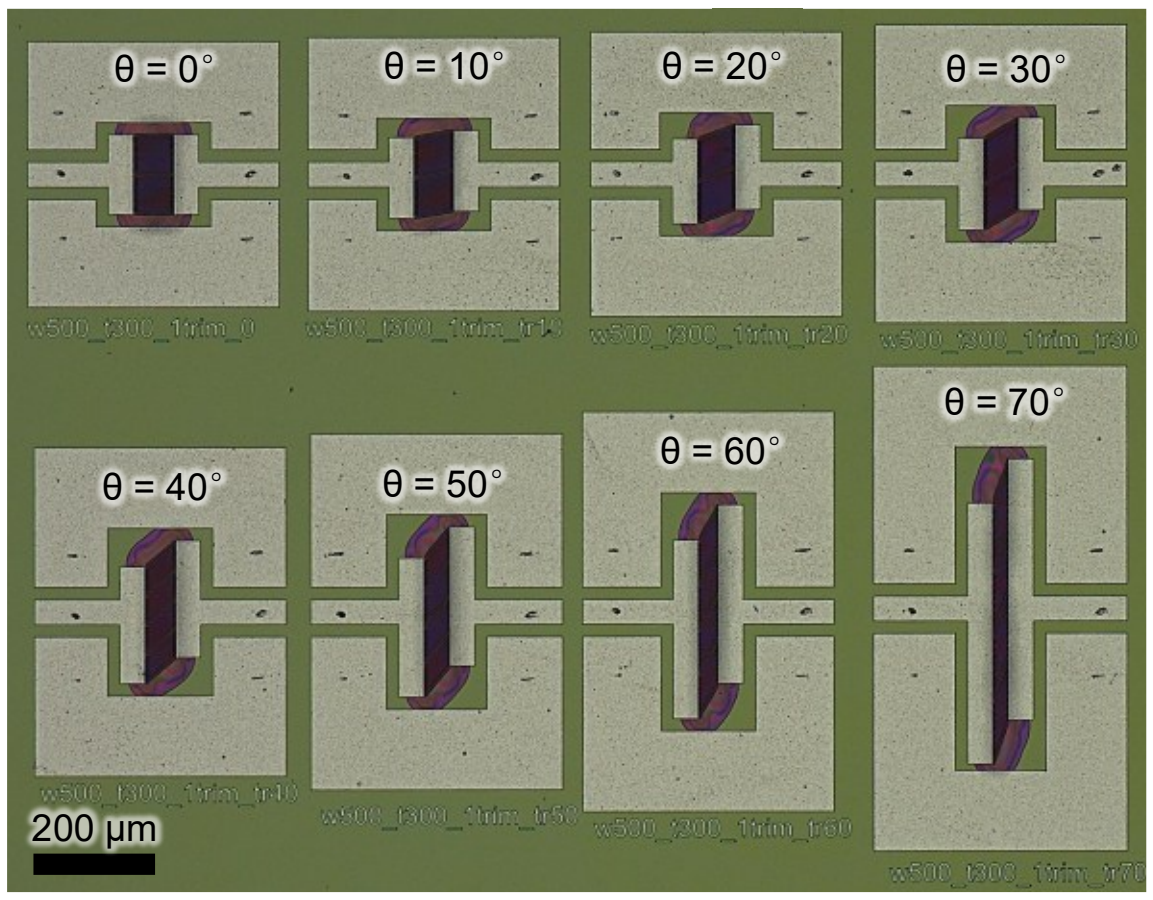

Fig. 10. A group of fabricated resonators with varying IDE angles, but identical number of trims.

Following the thinning step, aluminum metallization is patterned on the trimmed wafer using electron beam lithography (EBL). The metallization consists of 350 nm-thick IDEs and 700 nm-thick buslines. After the Al deposition, etch windows are defined by ion milling. These windows serve to confine the active regions and enable thin-film suspension. The release process is performed using xenon difluoride ($XeF_2$), which selectively removes the sacrificial amorphous silicon (aSi) layer beneath the TFLN. This step must be executed with care, as the small-featured IDEs are the sole structural support for the suspended resonators and are susceptible to collapse if over-etched or poorly supported.

In this work, we fabricated and characterized three sets of test structures to validate the proposed design methodology: individual resonator behavior, and two prototype filter designs. First, to verify the concept of resonator rotation, two study groups of XBARs with varying in-plane rotation angles and film thicknesses were fabricated at two different locations. For the prototype filters, their constituent resonators were also fabricated separately alongside the filter structures to enable direct monitoring of their individual resonance responses.

### B. Oriented Resonator Measurements

Two study groups, each comprising three XBAR sets with different TFLN thicknesses, were fabricated at separate wafer locations and measured. Each set contained eight devices with IDE tilting angles ranging from 0° to 70°, as shown in Fig. 10. All measurements were performed using a Keysight E8361C PNA vector network analyzer (VNA) with GSG probes.

The series resonant frequency $f_s$ of each resonator was extracted from the measured two-port $S$-parameters, via standard $Y_{21}$ conversion [9]. As shown in Fig. 11(a), three distinct $f_s$ ranges are observed, corresponding to the three TFLN thicknesses. Comparison with simulation predictions, extrapolated from $f_s$ at θ = 0°, the effects of local thickness variation are evident. A similar trend is observed in Fig. 11(b), though with shifted frequencies, indicating global thickness variations across the same wafer.

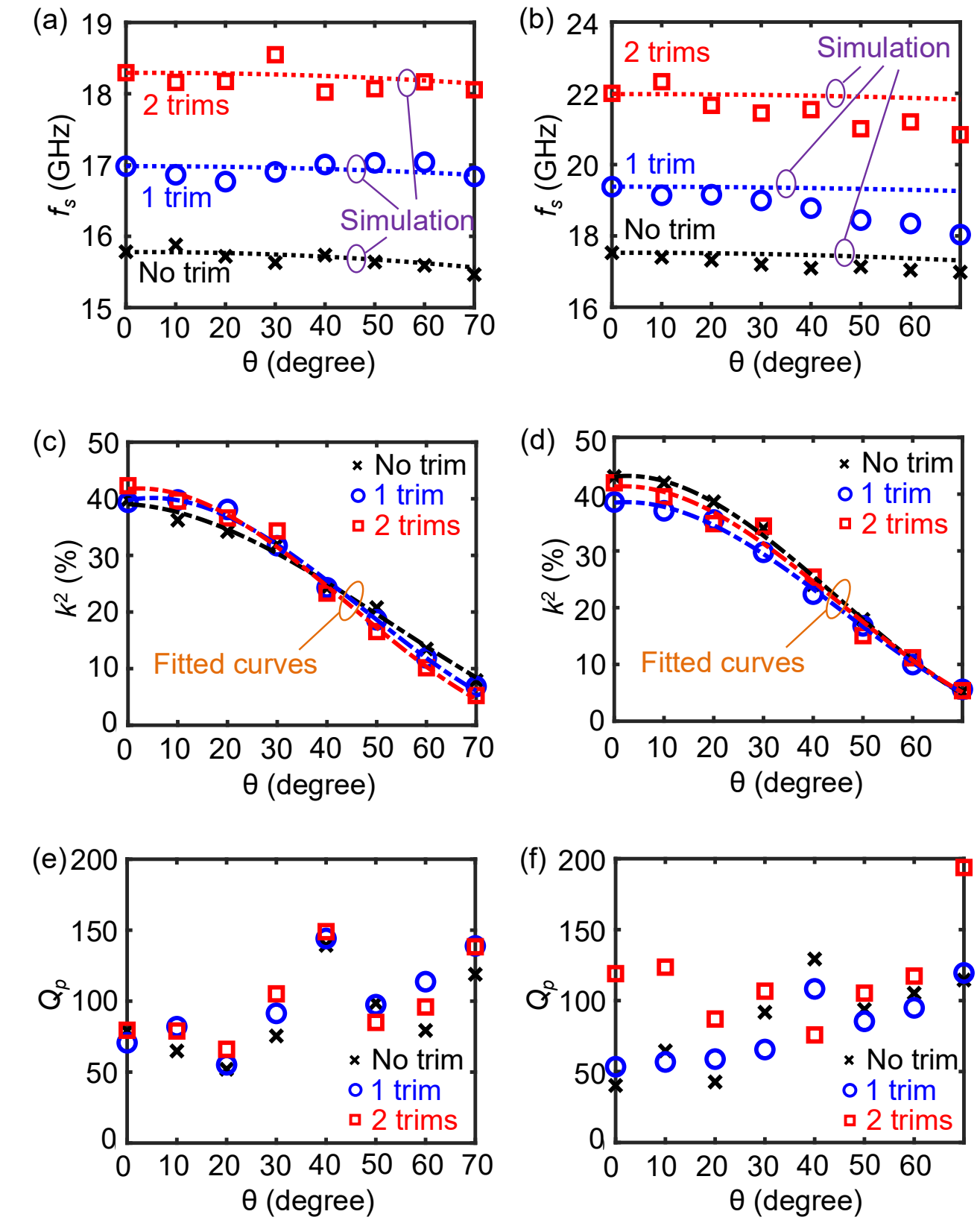

Fig. 11. Extracted resonator parameters: (a) and (b) $f_s$ vs. θ with simulation-based predictions (dotted lines), (c) and (d) $k^2$ vs. θ with fitted curves (dashed lines), and (e) and (f) $Q_p$ vs. θ from study sets 1 and 2, respectively.

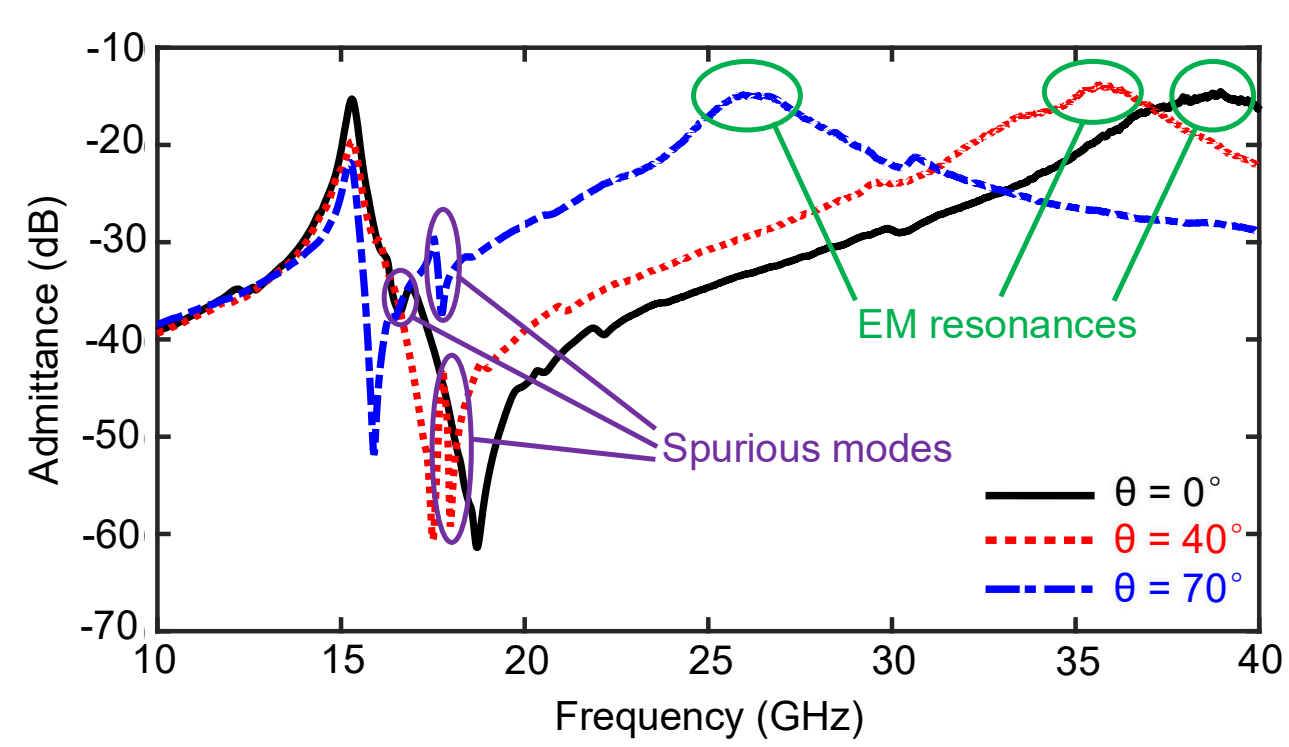

Fig. 12. Measurement samples from the untrimmed group of the study Set 1, indicating spurious modes and EM resonance.

Figs. 11(c) and (d) present the extracted $k^2$, along with fitted curves from the two study groups. The results align well with the simulations in Fig. 4, with minor discrepancies attributed to spurious modes, as well as parasitic inductance $L_s$ due to EM resonance, as illustrated in Fig. 12. These effects introduce inaccuracies when directly apply (4) to determine $k^2$ from the measured $f_s$ and $f_p$. Accurate extraction of $k^2$ therefore require modeling based on the mBVD circuit [61]. For detailed discussions on precise modeling of detailed discussions on modeling piezoelectric devices under EM interaction, readers are referred to [62].

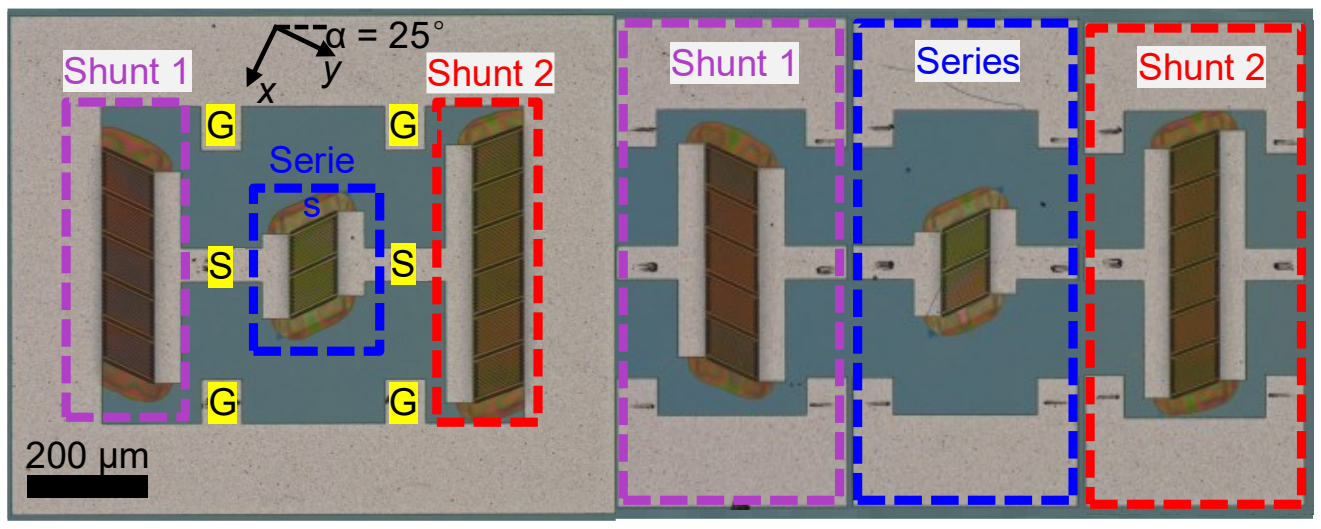

Fig. 13. The fabricated three-element filter prototype and its standalone constituent resonators.

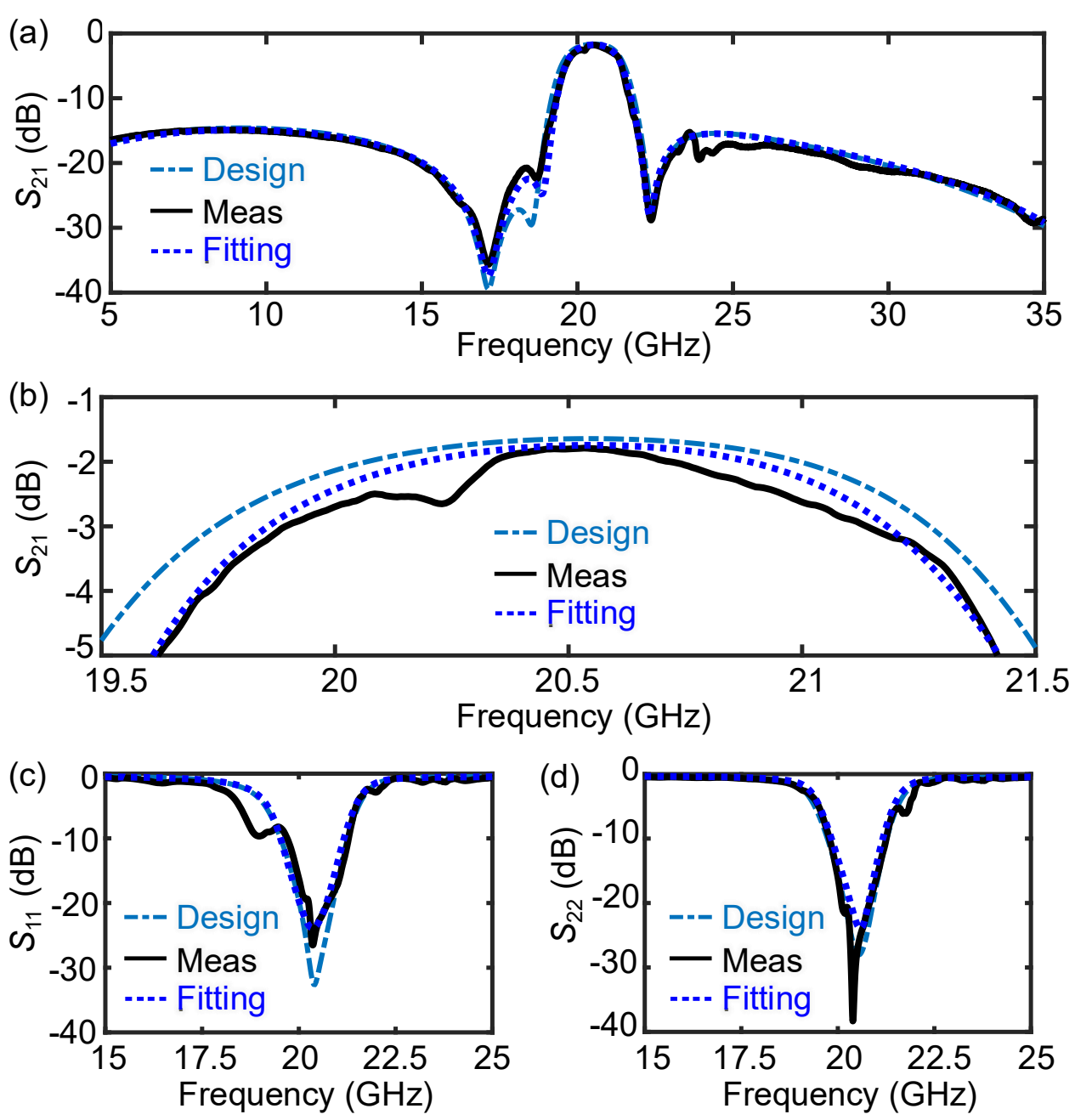

Fig. 14. Measurement results compared to the simulation-based design and mBVD post-measurement fitting of the three-element prototype: (a) $S_{21}$, (b) a magnified passband, (c) $S_{11}$, and (d) $S_{22}$.

TABLE III
SIMULATION AND MEASUREMENT COMPARISONS FOR THREE-ELEMENT FILTER CHARACTERISTICS

| Parameters | Simulation | Measurement | Fitted model |
|---|---|---|---|
| $f_c$ | 20.5 GHz | 20.5 GHz | 20.5 GHz |
| Min. IL | 1.69 dB | 1.79 dB | 1.79 dB |
| 3dB-FBW | 9.54% | 8.58% | 8.78% |
| 20dB-FBW | 16.8% | 16.5% | 15.9% |
| Lower OoB Rej. | 14.59 dB @9.2 GHz | 14.92 dB @8.6 GHz | 14.78 dB @9.7 GHz |
| Upper OoB Rej. | 15.42 dB @24.5 GHz | 15.24 dB @23.6 GHz | 16.36 dB @24.8 GHz |

Regarding the quality factor of the rotated XBARs, $Q_p$ is considered since it is less affected by EM resonances. As shown in Figs. 11(e) and (f), the extracted $Q_p$ from both study groups exhibits insignificant dependence on orientation angle, ranging from 50 to 200. The results suggest that increasing θ tends to improve $Q_p$ as spurious modes are suppressed by lower $k^2$; however, comprehensive investigation is left for future work.

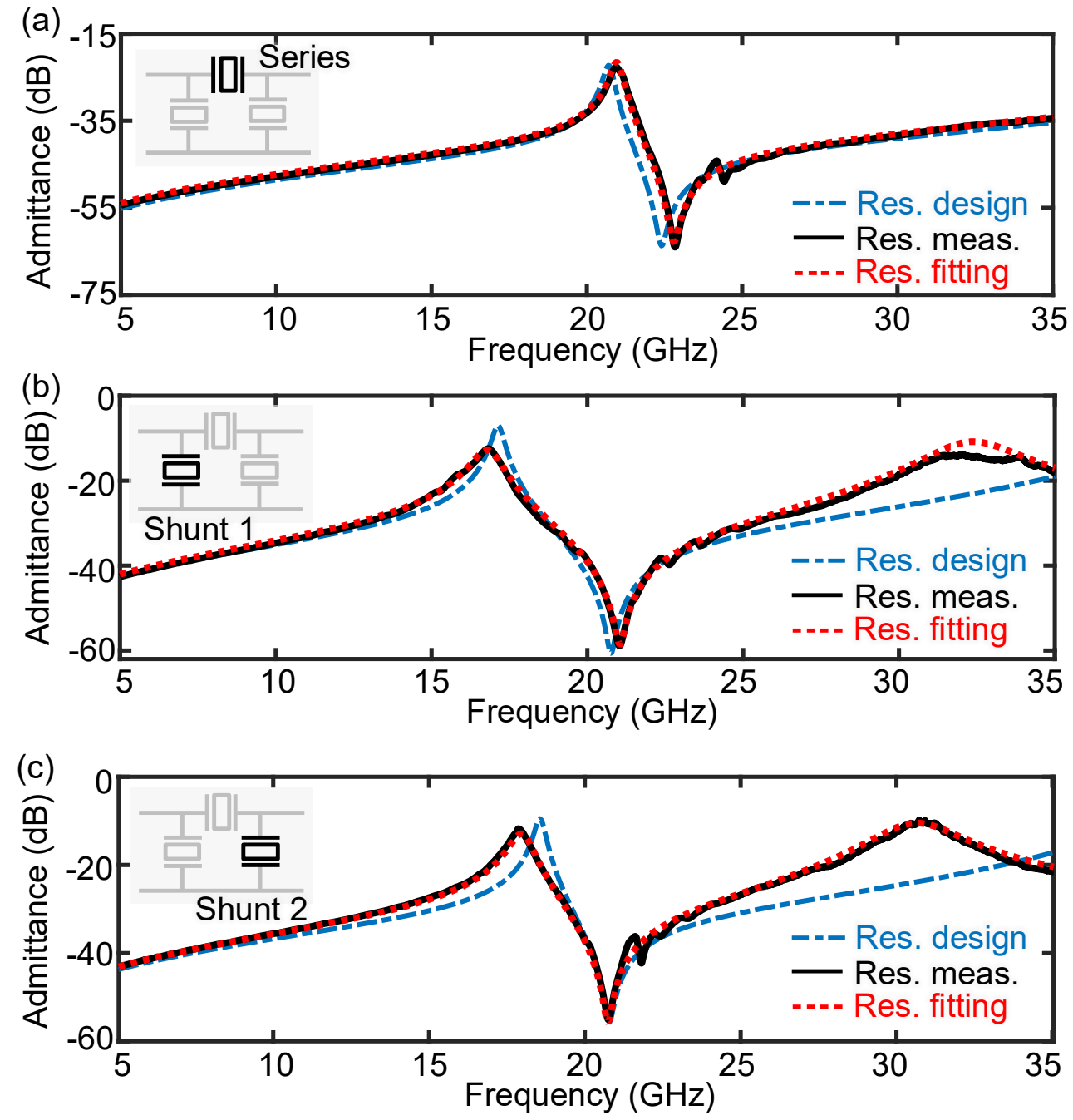

Fig. 15. Admittance comparisons of the design, individual resonator measurement, and resonator fitting of the three-element prototype: (a) Series, (b) Shunt 1, and (c) Shunt 2, respectively.

TABLE IV
MBVD MODEL PARAMETER COMPARISONS OF THREE-ELEMENT FILTER

| Resonator | | $f_s$ (GHz) | $k^2$ (%) | $Q$ | $C_0$ (fF) | $R_s$ (Ω) | $L_s$ (nH) |
|---|---|---|---|---|---|---|---|
| Series | Design | 21.00 | 17.0 | 80 | 48 | 3 | 0.2 |
| | Filter fitting | 20.99 | 17.1 | 76 | 47 | 1.5 | 0.28 |
| | Res. fitting | 21.30 | 17.9 | 80 | 55 | 6 | 0.18 |
| Shunt 1 | Design | 17.90 | 42.5 | 80 | 168 | 1 | 0.1 |
| | Filter fitting | 18.05 | 41.5 | 76 | 177 | 1.5 | 0.11 |
| | Res. fitting | 18.25 | 40.7 | 70 | 178 | 3.9 | 0.16 |
| Shunt 2 | Design | 19.10 | 22.6 | 80 | 169 | 1 | 0.1 |
| | Filter fitting | 19.30 | 19.5 | 70 | 175 | 1.5 | 0.08 |
| | Res. fitting | 19.10 | 22.1 | 86 | 179 | 3.6 | 0.17 |

## C. Three-Element Filter Prototype

Fig. 13 shows the in-house fabricated three-element filter prototype alongside its standalone resonators. The measured $S_{21}$ obtained using the VNA is plotted in Fig. 14(a), together with the simulated response from the simulation-based design and the fitted post-processed mBVD network. We will use the fitted model to further analyze our design. A magnified view of the passband, shown in Fig. 14(b), demonstrates close agreement between measurement and simulation. Regarding impedance matching, Figs. 14(c) and (d) indicate that the filter is well-matched to a 50 Ω reference at both ports. The observed asymmetry is due to the use of different shunt resonators, which introduce different input impedances at each port.

Table III summarizes key filter metrics obtained from the simulation-based design, measurement, and the fitted model for the three-element filter prototype. Note that the OoB rejection in this work is defined by a minimum IL within the frequency range $f_c$ ± 10BW from the passband. The fabricated device exhibits a slightly higher insertion loss of 1.79 dB and a reduced FBW of 8.58%, compared to the designed 9.54%. The

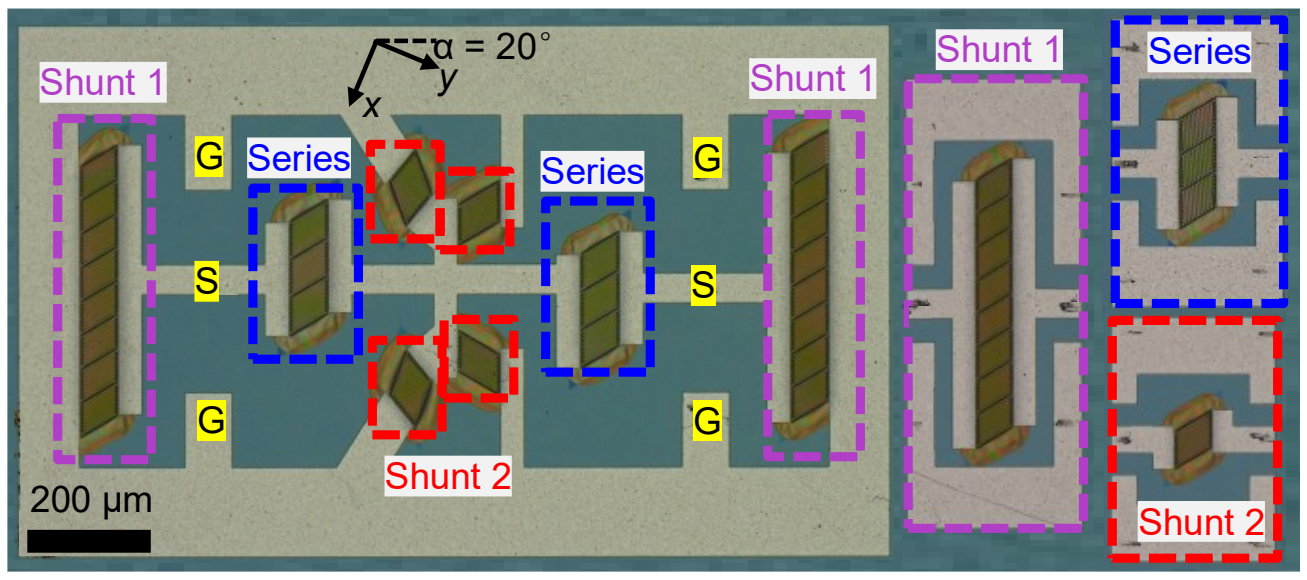

Fig. 16. The fabricated eight-element filter prototype and its standalone constituent resonators.

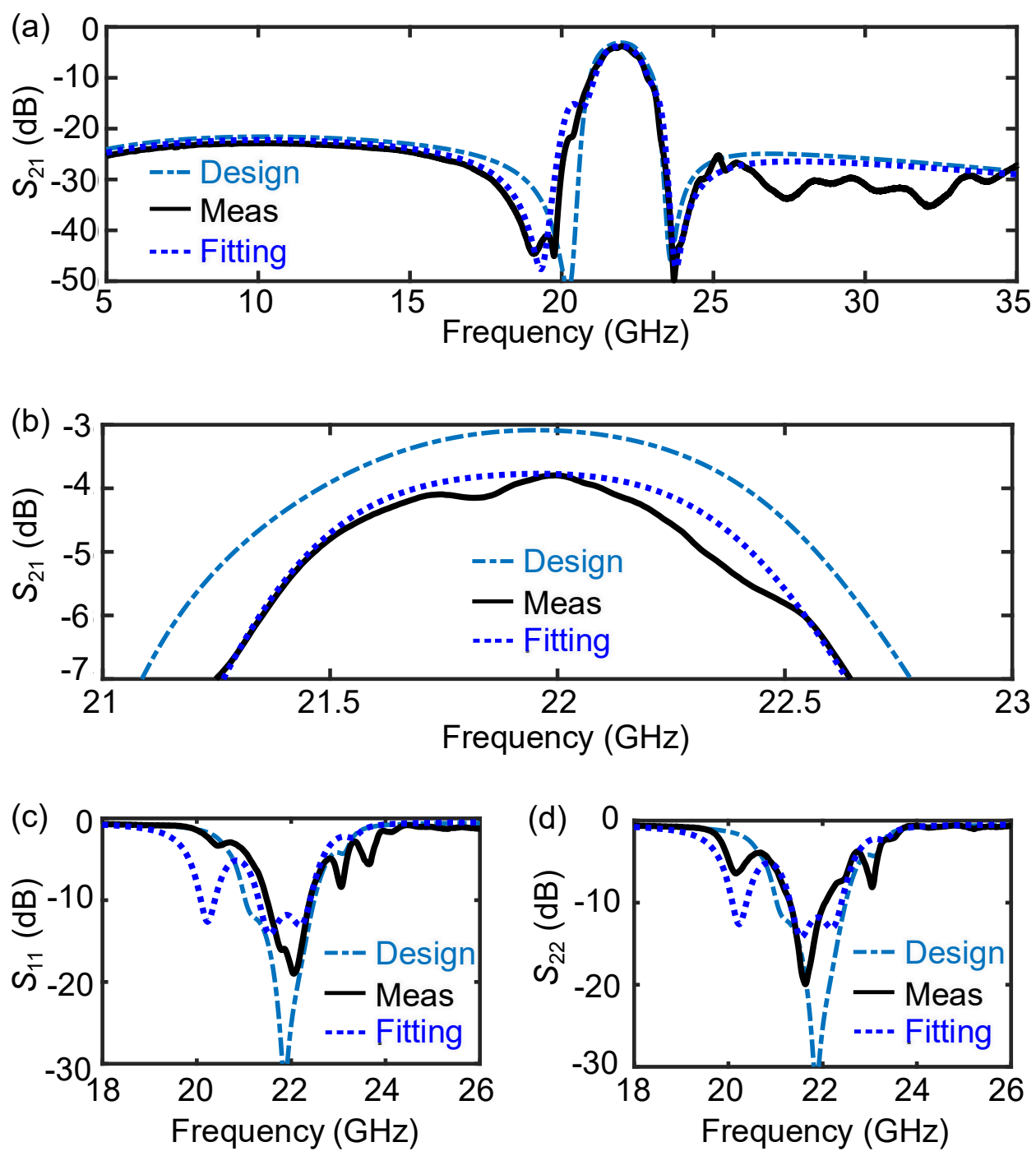

Fig. 17. Measurement results compared to the simulation-based design and mBVD post-measurement fitting of the eight-element prototype: (a) $S_{21}$, (b) a magnified passband, (c) $S_{11}$, and (d) $S_{22}$.

TABLE V
SIMULATION AND MEASUREMENT COMPARISONS FOR THREE-ELEMENT FILTER CHARACTERISTICS

| Parameters | Simulation | Measurement | Fitted model |
|---|---|---|---|
| $f_c$ | 22.0 GHz | 22.0 GHz | 22.0 GHz |
| Min. IL | 3.08 dB | 3.80 dB | 3.80 dB |
| 3dB-FBW | 6.40% | 6.12% | 6.12% |
| 20dB-FBW | 12.1% | 14.5% | 18.8% |
| Lower OoB Rej. | 21.57 dB @10.4 GHz | 22.97 dB @11.5 GHz | 22.24 dB @10.2 GHz |
| Upper OoB Rej. | 25.94 dB @26.8 GHz | 25.30 dB @25.2 GHz | 26.45 dB @27.8 GHz |

measured admittance responses of standalone resonators fabricated adjacent to the filter are compared with the design values, as shown in Figs. 15(a)–(c), to examine the discrepancies. According to Table IV, the fitted mBVD model parameters extracted from the fabricated filter closely align with the design values, indicating only minor deviations and contributing to the overall accurate response. It can be inferred that a primary source of deviation is the slightly overestimated $k^2$ in the simulation, which significantly impacts the FBW. In addition, layout-induced parasitics, $R_s$ and $L_s$, are other

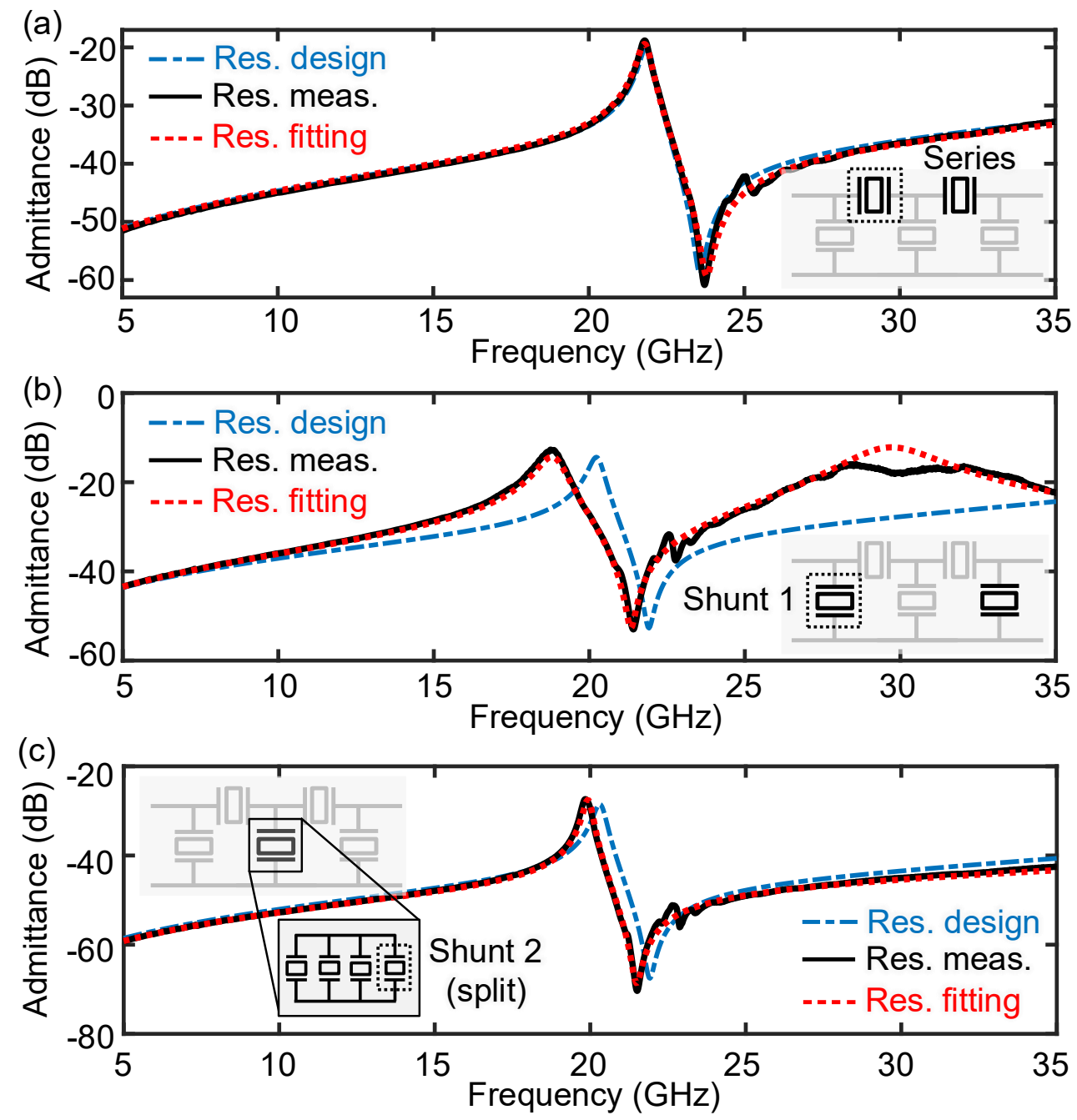

Fig. 18. Admittance comparisons of the design, individual resonator measurement, and resonator fitting of the eight-element prototype: (a) Series, (b) Shunt 1, and (c) Shunt 2, respectively.

TABLE VI
MBVD MODEL PARAMETER COMPARISONS OF EIGHT-ELEMENT FILTER

| Resonator | | $f_s$ (GHz) | $k^2$ (%) | $Q$ | $C_0$ (fF) | $R_s$ (Ω) | $L_s$ (nH) |
|---|---|---|---|---|---|---|---|
| Series | Design | 22.13 | 16.5 | 80 | 77 | 3.5 | 0.2 |
| | Filter fitting | 22.10 | 19.5 | 70 | 71 | 3.2 | 0.28 |
| | Res. fitting | 22.10 | 17.5 | 70 | 75 | 3 | 0.18 |
| Shunt 1 | Design | 20.50 | 18.0 | 80 | 180 | 2.5 | 0.1 |
| | Filter fitting | 19.49 | 18.8 | 60 | 182 | 2.0 | 0.11 |
| | Res. fitting | 20.00 | 17.0 | 80 | 180 | 4 | 0.16 |
| Shunt 2# | Design | 20.50 | 17.5 | 80 | 32.5 | 2.5 | 0.1 |
| | Filter fitting | 20.83 | 19.0 | 70 | 31 | 5.0 | 0.08 |
| | Res. Fitting | 20.00 | 19.0 | 76 | 30 | 2.8 | 0.17 |

#Split resonator.

contributing factors. Furthermore, spurious modes present in the fabricated prototype potentially contribute to the passband ripples observed in the filter measurement.

## D. Eight-Element Filter Prototype

The images of the fabricated eight-element filter prototype and its standalone resonators are given in Fig. 16. Compared to the three-element design, this prototype poses greater challenges for modeling with a simple mBVD circuit network (see Fig. 17(a)), due to increased complexity from multiple resonator variations, a larger layout, and stronger spurious mode effects. To reduce complexity in fitting the filter response, we assume that the two Shunt 1 resonators are identical, as are the four Shunt 2 resonators, in line with the original design intent. Despite this simplification, the comparisons in Table V and the close-up view in Fig. 17(b) confirm that the proposed modeling approach is sufficiently accurate for passband characterization. Notably, the asymmetry in the S-parameters, as shown in Figs. 17(c) and (d), can be attributed to several factors, including mismatches between resonators that were designed to be identical and the influence of parasitic EM effects.

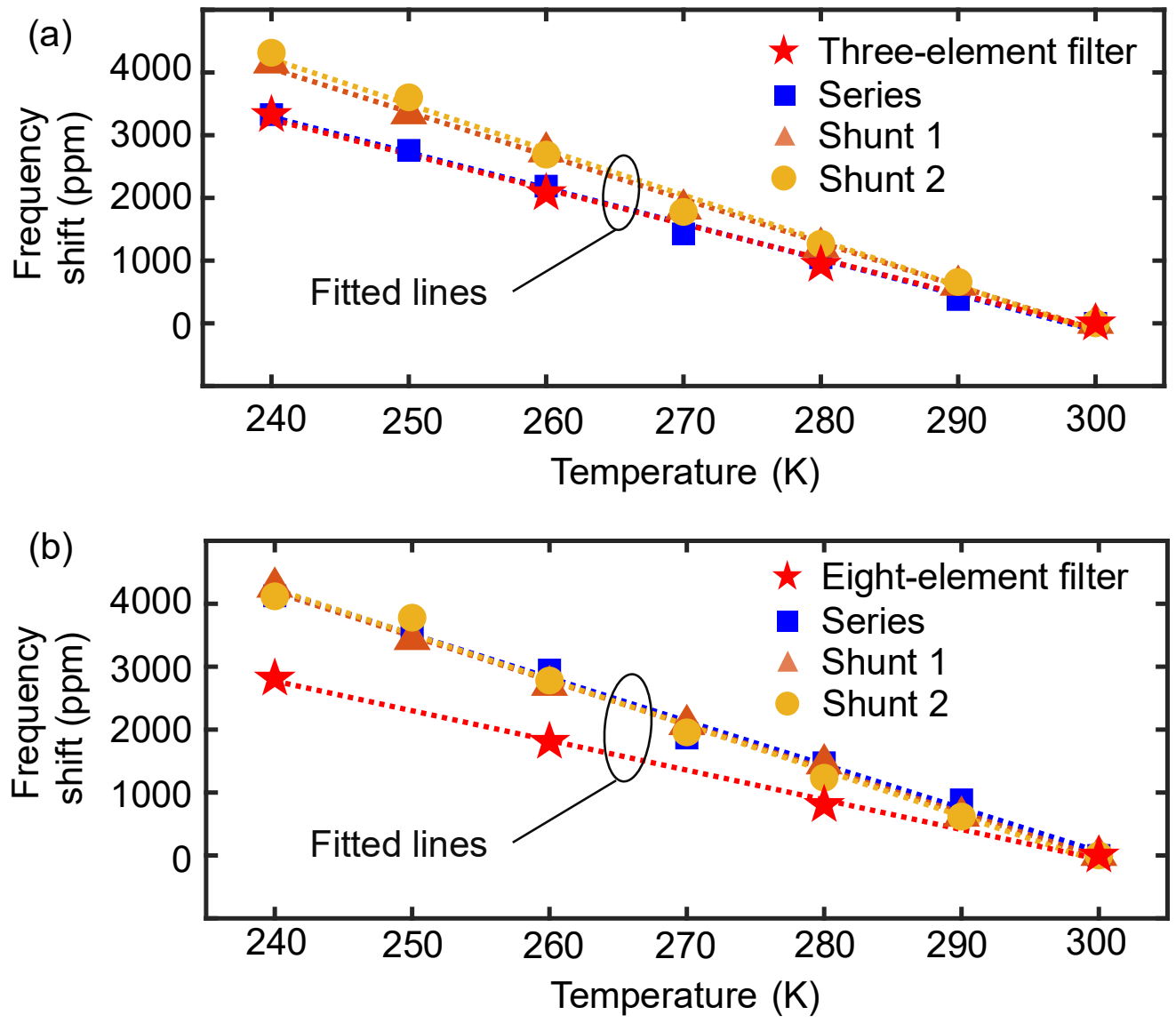

Fig. 19. Measured and fitted (dotted lines) TCF measurements of (a) the three-element prototype and its resonators and (b) the eight-element prototype and its resonators.

To further investigate the discrepancies, the standalone resonator measurements and their corresponding mBVD model fittings are compared with the design parameters, as shown in Figs. 18(a)–(c). Following the data summarized in Table VI, the design and measured parameters, both from filter fitting and standalone resonator fitting, are generally well-aligned, with minor discrepancies in $f_s$ and $k^2$ due to thickness variations and simulation accuracy. It is important to note that $R_s$ and $L_s$ are inherently layout-dependent and are expected to differ across the three cases. These results suggest that the main contributors to the observed deviations are likely mismatches between nominally identical resonators, as well as the influence of EM effects.

### E. Temperature Dependency of FR3 XBAR Devices

The temperature coefficient of frequency (TCF) of the fabricated A1 XBAR devices was evaluated by measuring their S-parameters at different temperatures range from 240 K to 300 K in a high vacuum environment with pressure of below $10^{-5}$ Torr. The measurement setup, similar to that in [63], consists of a Lakeshore CRX-6.5K probe station, connected to the VNA and a liquid helium cooling system for precise temperature control. TCF is defined as

$$\mathrm{TCF} = \frac{1}{f_0}\frac{\Delta f}{\Delta T}, \quad (5)$$

where $\Delta f$ and $\Delta T$ denote the frequency and temperature differences relative to the reference at $T = 300$ K in this study. For the individual resonators, $f$ corresponds to $f_p$ in this study to mitigate the impact of EM resonances, whereas $f_c$ is considered in the case of filters.

The normalized frequency shifts $\Delta f/f_0$ extracted from the measurements down to 240 K, with step sizes of 20 K for the two filter prototypes and 10 K for their resonators, are plotted in Fig. 19. Linear regression of the data indicates that both $f_p$ and $f_c$ of the A1 XBAR resonators and filters decrease linearly with temperature. Table VII presents TCFs calculated from the fitting, along with the coefficient of determination $R^2$, of the measured devices. Despite a poor TCF of about –70 ppm/K of the TFLN platform, which clearly reflects through individual resonator TCF performance, a combination of them surprisingly provides improved TCF values of –55.53 ppm/K and –47.20 ppm/K in the three- and eight-element filter prototypes, without any engineered TCF compensation. Notably, larger devices tend to improve TCF, especially in the case of filters with more resonators. It is essential to note that a number of factors such as crystal orientation, incorporated materials (e.g., substrate and metallization), and device structure can influence the TCF of acoustic devices. Further systematic investigations will be carried out to improve understanding and explore potential improvements to this platform.

TABLE VII
TCF FITTING SUMMARY

| A1 XBAR Devices | | TCF (ppm/K) | $R^2$ |
|---|---|---|---|
| Three-element filter | Filter | –55.53 | 0.996 |
| | Series | –56.72 | 0.995 |
| | Shunt 1 | –69.43 | 0.996 |
| | Shunt 2 | –72.34 | 0.992 |
| Eight-element filter | Filter | –47.20 | 0.997 |
| | Series | –69.11 | 0.990 |
| | Shunt 1 | –70.19 | 0.998 |
| | Shunt 2* | –72.22 | 0.992 |

*Split resonator.

### F. Discussions and Outlook

Our prototype fabrication and measurements have highlighted two important intertwined factors essential for designing acoustic BPF with precise control over the FBW and $f_c$: modeling accuracy and process variations. While $f_c$ can be estimated from $f_s$ of the series resonator, which is largely determined by the precision of active region etching and the thickness uniformity of the TFLN platform, it is evident that accurate prediction of $k^2$ is most crucial. The value of $k^2$ strongly influences overall filter performance, as it governs the location of TZs and, consequently, underpins the required etching depths of all active regions from the initial wafer thickness. The findings show that even small variations in resonant frequencies can result in noticeable discrepancies, especially in the higher-order BPF containing multiple resonators. It is worth mentioning that such film thickness variations have more pronounced effects in multilayer periodically poled piezoelectric film (P3F) platforms [42]–[44], [64], [65], where thickness imbalances can significantly degrade the performance and introduce additional spurious modes. Notably, the same thickness trimming and resonator orientation techniques demonstrated in this work can also be applied to a P3F 128° Y-cut TFLN platform [59]. The P3F platform is expected to further enhance XBAR performance, and with experimental validation left for future work. Nevertheless, stringent fabrication quality control will be essential. Furthermore, future work will also study the loss and $Q$ dependency on the in-plane orientation for XBAR resonators, as well as thermal and nonlinear behaviors,

TABLE VIII
REPORTED FR3 ACOUSTIC FILTERS BEYOND 10 GHZ

| Ref. | Technology | $f_c$ (GHz) | IL (dB) | Rej. (dB)* | FBW (%) | Monolithic $f_c$ and FBW |
|---|---|---|---|---|---|---|
| [37] | TFLN XBAR | 23.5 | 2.3 | 13.6 | 18.2 | No |
| [38] | TFLN XBAR | 22.1 | 1.6 | 11.9 | 19.8 | No |
| [41] | TFLN XBAR with inductors | 19.0 | 8.0 | 13.0 | 2.4 | No |
| [42] | Bi-layer P3F AlScN FBAR | 10.7 | 0.7 | N/A | 5.3 | No |
| [43] | Bi-layer P3F TFLN XBAR | 23.8 | 1.5 | 9.5 | 19.4 | No |
| [44] | 4-layer P3F AlScN FBAR | 17.4 | 1.86 | 5.1 | 3.9 | No |
| | | 17.4 | 3.25 | 12.8 | 3.4 | No |
| **This work** | **TFLN XBAR** | **20.5** | **1.79** | **14.92** | **8.58** | **Yes** |
| | | **22.0** | **3.80** | **22.97** | **6.12** | **Yes** |

*OoB rejection is defined as the minimum IL within the frequency range $f_c \pm 10$BW from the passband and is extracted from the plot in the reference to maintain consistent comparisons; N/A: insufficient data.

TABLE IX
ADVANCED FILTER TECHNOLOGY COMPARISONS

| Ref. | Technology | $f_c$ (GHz) | Area ($\lambda_0^2$) | Height (mm) | IL (dB) | Rej. (dB)* | FBW (%) | Shape factor$^\$$ |
|---|---|---|---|---|---|---|---|---|
| [8] | GaAs-IPD | 28 | 0.003$^\dagger$ | N/A | 0.76 | 18.6 | 22 | 2.25 |
| [12] | DSG-PCB | 4.94 | 0.12 | 0.5 | 1.8 | 25.8 | 64 | 1.78 |
| [14] | SIW-PCB | 10.05 | 0.24 | 0.5 | 0.82 | 45 | 12.2 | 1.43 |
| [15] | SIW-TSV | 17.96 | 0.03 | 0.2 | 0.85 | N/A | 89.2 | 1.52 |
| [16] | SIW-TGV | 32 | 0.18 | 0.4 | 0.96 | N/A | 13.1 | 2.15 |
| [18] | SIW-LTCC | 27 | 0.001$^\dagger$ | 2.8$^{\dagger\dagger}$ | 1.36 | 22.0 | 18.2 | 3.38 |
| **This work** | **TFLN XBAR (MEMS)** | **20.5** | **0.003** | **0.5** | **1.79** | **14.92** | **8.58$^\#$** | **1.92** |
| | | **22.0** | **0.007** | **0.5** | **3.80** | **22.97** | **6.12$^\#$** | **2.37** |

* OoB rejection is defined as the minimum IL within the frequency range $f_c \pm 10$BW from the passband and is extracted from the plot in the reference to maintain consistent comparisons; $^\dagger$ excluded GSG probes; $^{\dagger\dagger}$ estimated from the reported layer thickness; $^\#$ customized FBW; $^\$$shape factor is the ratio of 20-dB FBW to 3-dB FBW; N/A: insufficient data.

reflecting through two-tone third-order intermodulation (IMD3), for better optimizing the resonator and filter design.

Although the simple mBVD circuit model is proven to offer sufficiently accurate predictions, the ability to account for electromagnetic (EM) resonances induced by the filter layout is indispensable. In fact, such EM behaviors could be strategically exploited to increase design flexibility through EM-acoustic co-design. This approach could enhance filter characteristics—for example, broadening the FBW by increasing the separation between $f_s$ and $f_p$ using higher-order ladder networks with added transmission zeros and modest EM loading via hybrid filter structures [41], [52], while independently improving out-of-band rejection through the strategic addition or repositioning of TZs. The efforts can extend beyond EM hybrids to include emerging microacoustic metamaterials [66] as well. Such directions require both appropriate filter synthesis and realizable topologies. Advancing along these directions will require both rigorous filter synthesis and practically realizable topologies. We plan to pursue and report FR3 hardware prototypes in future work.

As shown in Table VIII, our design approach has demonstrated significant advancements in acoustic filter design for FR3 operation in terms of both performance and design degrees of freedom. Our three-element prototype achieves low IL while retaining relatively high OoB rejection. Although the eight-element prototype incurs approximately 2 dB higher IL and exhibits impairments due to fabrication challenges, it provides an OoB rejection improvement of over 8 dB compared to its lower-order counterpart. More importantly, our innovation enables monolithic control over BPF characteristics, allowing for the implementation of multiple filters on a single wafer. With further refinement in device modeling and process accuracy, as well as considerations of novel monolithically switchable acoustic filter techniques [67], this approach could pave the way for high-performance, chip-scale monolithic BAW multiplexers and filter banks.

Compared to other advanced filter technologies, as indicated in Table IX, our work shows competitive overall performance, with a great balance between IL, customizable FBW, OoB rejection, and selectivity within an extremely compact planar footprint. A strong chip-scale competitor of acoustic technology at FR3 and beyond is IPD [8], which offers superior OoB rejection and significantly lower IL. However, our three-element filter achieves a sharper shape factor of 1.92, compared to 2.25 of the IPD counterpart, owing to the high-Q resonances of XBARs. Interestingly, the shape factor of the eight-element, expected to be of 1.73 according to our design, is instead slightly worse, likely due to the previously mentioned fabrication challenges associated with larger-scale structures. Nonetheless, continued improvements in design and implementation could unleash the full potential of BAW filters for mobile applications operating in increasingly congested spectral environments.

## VI. CONCLUSION

This paper comprehensively introduces an approach for the design of $f_c$ and FBW in BAW filters at FR3 by exploiting in-plane anisotropic properties of 128° Y-cut TFLN, enabling flexible filter customization solely through planar orientation and local film thickness control. The design demonstration—from multiphysics device simulation to fabrication—culminates in two BPF prototypes with tailored FBWs. A thorough analysis reveals key insights and challenges associated with modeling and fabrication. With further development, the proposed methodology would pave the way for chip-scale multiplexers and filter banks with precise band selection for next-generation wireless mobile communications.

## ACKNOWLEDGEMENT

The authors thank Dr. Ben Griffin, Dr. Todd Bauer, and Dr. Zachary Fishman for helpful discussions.